\documentclass{aastex62}

\usepackage[figuresright]{rotating}

\shorttitle{Droplets II}
\shortauthors{Chen et al.}

\begin{document}

\title{Droplets II: Internal Velocity Structures and Potential Rotational Motions in Pressure-dominated Coherent Structures}

\correspondingauthor{Hope How-Huan Chen}
\email{hopechen@utexas.edu}

\author[0000-0001-6222-1712]{Hope How-Huan Chen}
\affil{Department of Astronomy, The University of Texas, Austin, TX 78712, USA}
%\affil{Harvard-Smithsonian Center for Astrophysics, 60 Garden St., Cambridge, MA 02138, USA}
%\collaboration{(Collaboration)}

\author{Jaime E. Pineda}
\affil{Max-Planck-Institut f\"ur extraterrestrische Physik, Giesenbachstrasse 1, D-85748 Garching, Germany}

\author{Stella S. R. Offner}
\affil{Department of Astronomy, The University of Texas, Austin, TX 78712, USA}

\author{Alyssa A. Goodman}
\affil{Harvard-Smithsonian Center for Astrophysics, 60 Garden St., Cambridge, MA 02138, USA}

\author{Andreas Burkert}
\affil{University Observatory Munich (USM), Scheinerstrasse 1, 81679 Munich, Germany}

\author{Rachel K. Friesen}
\affil{Department of Astronomy \& Astrophysics, University of Toronto, 50 St.\ George St., Toronto, ON M5S 3H4, Canada}
\affil{National Radio Astronomy Observatory, 520 Edgemont Rd., Charlottesville, VA, 22903, USA}

\author{Erik Rosolowsky}
\affil{Department of Physics, 4-181 CCIS, University of Alberta, Edmonton, AB T6G 2E1, Canada}

\author{Samantha Scibelli}
\affil{Steward Observatory, 933 North Cherry Ave., Tucson, AZ 85721, USA}

\author{Yancy Shirley}
\affil{Steward Observatory, 933 North Cherry Ave., Tucson, AZ 85721, USA}

\begin{abstract}
We present an analysis of the internal velocity structures of the newly identified sub-0.1 pc coherent structures, \textit{droplets}, in L1688 and B18.  By fitting 2D linear velocity fields to the observed maps of velocity centroids, we determine the magnitudes of linear velocity gradients and examine the potential rotational motions that could lead to the observed velocity gradients.  The results show that the droplets follow the same power-law relation between the velocity gradient and size found for larger-scale dense cores.  Assuming that rotational motion giving rise to the observed velocity gradient in each core is a solid-body rotation of a rotating body with a uniform density, we derive the ``net rotational motions'' of the droplets.  We find a ratio between rotational and gravitational energies, $\beta$, of $\sim 0.046$ for the droplets, and when including both droplets and larger-scale dense cores, we find $\beta \sim 0.039$.  We then examine the alignment between the velocity gradient and the major axis of each droplet, using methods adapted from the histogram of relative orientations (HRO) introduced by \citet{Soler_2013}.  We find no definitive correlation between the directions of velocity gradients and the elongations of the cores.  Lastly, we discuss physical processes other than rotation that may give rise to the observed velocity field.
\end{abstract}
% 0.046 and 0.039

%\keywords{ISM: clouds --- ISM: kinematics and dynamics --- ISM: structure --- stars: formation --- radio lines: ISM --- ISM: individual (L1688, B18)}

\section{Introduction}
\label{sec:intro}
\citet{Shu_1987} examined analytical star formation models and summarized an evolutionary sequence of a slowly rotating core with accretion initiated by an inside-out gravitational collapse.  Since then, it has been deemed important to characterize the rotational motion of the initial dense core that sets the stage for inside-out collapse and the formation of a deeply embedded disk.  At the time when the paper by \citet{Shu_1987} was first published, however, most systematic observational attempts to measure rotational motions in molecular clouds were made primarily based on analyses of $^{13}$CO observations, focusing on the more extended and less dense cloud material surrounding potentially star-forming cores \citep{Arquilla_1984, Arquilla_1985, Goldsmith_1985, Arquilla_1986}.  

\citet{Goodman_1993} (hereafter G93) summarized observations of denser gas, using NH$_3$ emission in 43 cores \citep[from][]{Benson_1989, Ladd_1994} and presented the first comprehensive work on measuring velocity gradients as a means to estimate angular momentum in dense cores.  Since then, that work, now nearly a quarter-century old, has become the standard reference for angular momentum and rotational energy values input to models of star and planet formation \citep[e.g.][]{Allen_2003, Li_2011, Seifried_2011}.  Meanwhile, our view of dense cores and star forming regions has changed significantly, both observationally and theoretically.  Extended analyses of observations and simulations have shown that cores are situated in the densest parts of a network of filamentary structures, often seen at an intersection of filaments \citep{McKee_2007, Myers_2009, Arzoumanian_2013}.  Filamentary structures are also shown to host most of the star forming cores \citep{Andre_2014, Padoan_2014, Hacar_2013, Tafalla_2015, Monsch_2018}.

Using observations of OH and C$^{18}$O line emission, \citet{Goodman_1998} proposed a characteristic size scale of $\sim$ 0.1 pc within which the scaling law between the linewidth and the size changes, from a power-law to a constant, nearly thermal value \citep[a \textit{Type 4} linewidth-size relation; see Fig.\ 9 in][]{Goodman_1998}.  Using GBT observations of NH$_3$ hyperfine line emission, \citet{Pineda_2010} made the first direct observation of a \textit{coherent core}, defined by a boundary across which the observed velocity dispersion changes from a turbulent regime to a coherent and nearly thermal one.
% and within the characteristic size scale from the center of a dense core, the linewidth is nearly thermal and virtually constant

Most recently, \citet{Chen_2019} (hereafter Paper I) used data from the Green Bank Ammonia Survey \citep[GAS;][]{GAS_DR1} and identified a total of 18 sub-0.1 pc coherent structures in L1688 (in Ophiuchus) and in B18 (in Taurus).  Each of the 18 coherent structures is identified by a change in velocity dispersion from a supersonically turbulent regime to a ``coherent''---uniform and subsonic---regime across the boundary, within which a centrally condensed distribution of NH$_3$ emission that traces the cold, dense gas usually associated with a dense core is found.  These coherent structures have a median size (radius) of $\sim$ 0.04 pc and a median mass of $\sim$ 0.4 M$_\sun$ and are termed ``\textit{droplets}'' owing to their small masses and sizes.  Paper I finds that the droplets appear to be an extension of the population of previously known larger-scale coherent cores \citep{Goodman_1993, Pineda_2010} at a smaller size scale, following the same core-to-core power-law mass-size and linewidth-size relations.  The droplets are also shown to be mostly virially unbound by self-gravity and primarily confined by the pressure provided by the ambient gas motions.

%%%%%% A paragraph here describing the turbulent models (Offner et al 2008; Dib et al. 2010; Chen & Ostriker).  In most of these models, the rotational motions are found not to be solid-body.

%% halve the following paragraph; it's more like an abstract with results now.
In this paper, we present a systematic analysis of the velocity gradients found in the 18 sub-0.1 pc coherent structures---droplets---in L1688 and in B18 identified in Paper I, and compare the results to the analysis of larger-scale dense cores presented by G93, with the goal of providing an updated standard of angular momentum and rotational energy measurements.  In \S\ref{sec:data}, we describe our data, including data from the GAS DR1 \citep[\S\ref{sec:data_GAS};][]{GAS_DR1}, maps of column density and dust temperature based on SED fitting of observations made by the Herschel Gould Belt Survey \citep[\S\ref{sec:data_Herschel};][]{Andre_2010}, and the catalogues of the droplets (Paper I) and larger-scale NH$_3$ cores \citep[\S\ref{sec:data_catalogs};][]{Goodman_1993}.

In \S\ref{sec:analysis_gradient}, we estimate the velocity gradients in droplets by fitting the observed velocity centroids in the local standard of rest velocity ($V_\mathrm{LSR}$) around each droplet with a 2D linear velocity field.  We then analyze the potential rotational motions that can give rise to the fitted velocity gradient, for the 13 droplets where the linear velocity fits produce reliable measurements of velocity gradients.  In \S\ref{sec:analysis_JoverM}, we derive the specific angular momentum and find that droplets appear to follow a power-law relation between the specific angular momentum and the size similar to the relation found for larger-scale dense cores by G93.  In \S\ref{sec:analysis_beta}, we derive the rotational energy and compare it to the gravitational energy and the kinetic energy derived in Paper I.  We then calculate the ratio between the rotational and gravitational energies, $\beta$, as well as the ratio between rotational and kinetic energies and examine their relations with the size and the virial equilibrium of droplets and dense cores.  In \S\ref{sec:analysis_pixbypix}, we investigate the effects of assuming constant density and solid-body rotation in the calculation of rotational properties by using the observed column density in the derivation of the rotational properties.  We then examine the alignment between the velocity gradient and the droplet shape using a method adapted from the histogram of relative orientations, used by \citet{Soler_2013} and \citet{Planck_35} to quantify the alignment between polarization vectors and local column density gradients, in \S\ref{sec:analysis_shape}.  In \S\ref{sec:discussion_physics}, we discuss the physical interpretation of the results by comparing them with previous studies of rotational properties of structures found in simulations and observations.  We summarize the results in \S\ref{sec:conclusion}.

\section{Data}
\label{sec:data}

\subsection{Green Bank Ammonia Survey (GAS)}
\label{sec:data_GAS}
The Green Bank Ammonia Survey \citep[GAS;][]{GAS_DR1} is a Large Program at the \textit{Green Bank Telescope} (GBT) to map most ``Gould Belt'' star forming regions with A$_\mathrm{V}$ $\geq$ 7 mag visible from the northern hemisphere in emission from NH$_3$ and other key molecules.  The data used in this work are from the first data release (DR1) of GAS that includes four nearby star forming regions: L1688 in Ophiuchus, B18 in Taurus, NGC1333 in Perseus, and Orion A.  Here we use GAS observations covering L1688 and B18 to derive the velocity gradients of the droplets identified in Paper I.

L1688 in Ophiuchus sits at a distance of $137.3\pm6$ pc \citep{OrtizLeon_2017}, and B18 in Taurus sits at a distance of $135\pm20$ pc \citep{Schlafly_2014}.  At these distances, the GBT FWHM beam size at 23 GHz of 32\arcsec corresponds to $\sim$ 4350 AU (0.02 pc).  The GBT beam size at 23 GHz also matches well with the Herschel SPIRE 500 $\micron$ FWHM beam size of 36\arcsec \citep[see \S\ref{sec:data_Herschel} and discussions in][]{GAS_DR1}.  The distances used in this paper are consistent with the measurements made by \citet{Zucker_2019} using GAIA data.

\subsubsection{Fitting the NH$_3$ Line Profile}
\label{sec:data_GAS_fitting}
In the GAS DR1, a (single) Gaussian line shape is assumed in fitting spectra of NH$_3$ (1, 1) and (2, 2) hyperfine line emission \citep[see \S3.1][]{GAS_DR1}.  The fitting is carried out using the ``cold-ammonia'' model and a forward-modeling approach in the \texttt{PySpecKit} package \citep{pyspeckit}, which was developed by \citet{GAS_DR1} and built upon the results from \citet{Rosolowsky_2008a} and \citet{Friesen_2009} in the theoretical framework laid out by \citet{Mangum_2015}.  No fitting of multiple velocity components or non-Gaussian profiles was attempted in GAS DR1, but the single-component fitting produced good quality results in $\gtrsim$ 95\% of detections in all regions included in the GAS DR1.  From the fitting, we can obtain the velocity centroid and the velocity dispersion of emission along each line of sight, where we have sufficient signal-to-noise in NH$_3$ (1, 1) emission.  For lines of sight where we detect both NH$_3$ (1, 1) and (2, 2), the model described in \citet{GAS_DR1} provides estimates of parameters including the kinetic temperature and the NH$_3$ column density.

\subsection{Herschel Column Density Maps}
\label{sec:data_Herschel}
The \textit{Herschel} column density maps are derived from archival Herschel PACS 160 and SPIRE 250/350/500 $\micron$ observations of dust emission, observed as part of the Herschel Gould Belt Survey \citep[HGBS][]{Andre_2010}.  The zero point of emission at each wavelength is calibrated with \textit{Planck} observations of the same regions \citep{PlanckXI}.  The calibrated emission is first convolved to match the SPIRE 500 $\micron$ beam FWHM of 36\arcsec and then passed to a least squares fitting routine, where the emission at these wavelengths is assumed to follow a modified blackbody emission function, $I_\nu = \tau B_\nu(T)$, where $B_\nu(T)$ is the blackbody radiation, and $\tau$ is the opacity.  The opacity can be written as a function of the mass column density, $\tau = \kappa_\nu \Sigma$, where $\kappa_\nu$ is defined as the opacity coefficient.  At these wavelengths, $\kappa_\nu$ can be described by a power-law function of frequency, $\kappa_\nu = \kappa_{\nu_0} \left(\frac{\nu}{\nu_0}\right)^\beta$, where $\beta$ is the emissivity index, and $\kappa_{\nu_0}$ is the opacity coefficient at the frequency, $\nu_0$.  Here we adopt $\kappa_{\nu_0}$ of 0.1 cm$^2$ g$^{-1}$ at $\nu_0$ = 1000 GHz \citep{Hildebrand_1983} and a fixed $\beta$ of 1.62 \citep{PlanckXI}.  The resulting $I_\nu$ is a function of the temperature and the mass column density, the latter of which can be further converted to the number column density by defining an average molecular weight \citep[2.8 u in this paper;][]{Kauffmann_2008}.  The resulting column density map has an angular resolution of 36\arcsec (the SPIRE 500 $\micron$ beam FWHM), which matches well with the GBT beam FWHM at 23 GHz (32\arcsec).  In the following analyses, no convolution is done to further match the resolutions of Herschel and GBT observations, before regridding the maps onto the same projection and gridding (Nyquist-sampled).

\subsection{Source Catalogs}
\label{sec:data_catalogs}
In this paper, we aim to provide a systematic analysis of rotational motions in dense cores by comparing the velocity gradients and other rotational properties found for the recently identified droplets (see \S\ref{sec:data_catalogs_droplets}; Paper I) to those found for larger-scale dense cores \citep[see \S\ref{sec:data_catalogs_Goodman93};][]{Goodman_1993}.  Below we describe the droplets and the cores.

% move this to a summary as the first section in analysis?
\subsubsection{Droplets from Paper I}
\label{sec:data_catalogs_droplets}
In Paper I, \citet{Chen_2019} identified a population of coherent structures, ``droplets,'' in L1688 and B18 using data from the GAS DR1.  The droplets are identified to be regions of subsonic velocity dispersion associated simultaneously with independent, NH$_3$-bright structures and with density structures on the \textit{Herschel} column density map.  In a virial analysis, the droplets are found to be mainly confined by the ambient gas pressure.  The radial density profiles of the droplets appear to be nearly constant at smaller radii.  At larger radii, the density profiles are steeper than but approaching $\rho \propto r^{-1}$.  Overall, the density profiles of the droplets are shallower than a Bonnor-Ebert sphere \citep[approaching $\rho \propto r^{-2}$ at large radii;][]{Ebert_1955,Bonnor_1956} and previous observations of starless cores \citep[e.g., $\rho \sim r^{-2.5}$ to $r^{-3.5}$ measured by][]{Tafalla_2004}.  See Fig.\ 12 in Paper I.

In Paper I, \citet{Chen_2019} also identified a set of coherent structures, ``droplet candidates,'' which are generally smaller regions with subsonic velocity dispersion.  Unlike droplets, the droplet candidates are neither independent, NH$_3$-bright structures nor associated with density structures on the \textit{Herschel} column density map.  In this work, we exclude the droplet candidates from any quantitative analysis and include them in the figures for reference.

\subsubsection{Dense Cores Measured in NH$_3$}
\label{sec:data_catalogs_Goodman93}
%%% This subsubsection has been checked and updated with AG's edits on Authorea.
G93 presented a survey of 43 sources with observations of NH$_3$ line emission (see Table 1 and Table 2 in G93; see also the \href{http://simbad.harvard.edu/simbad/sim-ref?querymethod=bib&simbo=on&submit=submit+bibcode&bibcode=1993ApJ...406..528G}{SIMBAD object list}).  For comparison with the kinematic properties of the droplets measured using observations of NH$_3$ hyperfine line emission from the GAS \citep{GAS_DR1}, we adopt values that were also measured using observations of NH$_3$ hyperfine line emission, presented by G93.  In this work, we adopt the updated physical properties presented in Paper I, which corrected the physical properties from G93 with the modern distance to each region.  The updated distances affect the physical properties listed in Table 1 in \citet{Goodman_1998}.  The size scales with the distance, $D$, by a linear relation, $R \propto D$.  Since the mass was calculated from the number density derived from NH$_3$ hyperfine line fitting, it scales with the volume of the structure, and thus $M \propto D^3$.  The updated distances also affect the velocity gradient, $\left|\mathcal{G}\right| \propto D^{-1}$, and other quantities derived from the velocity gradient in Table 2 in G93.  Besides the updated distances, the measurements of the kinetic temperature and the NH$_3$ linewidth, originally presented by \citet{Benson_1989} and \citet{Ladd_1994}, are used to derive the thermal and the non-thermal components of the velocity dispersion for the dense cores examined by G93.  See Paper I and \S\ref{sec:analysis_gradient} below for details.

% On the other hand, the coherent core observed by \citet{Pineda_2010} in the B5 region in Perseus is not included in the analysis of the net rotational motions in this paper, because it is not well-described by a linear velocity gradient.  

%\subsubsection{A Side Note: Coherent Core in B5}
%\label{sec:data_catalogs_Pineda10}
%Using GBT observations of NH$_3$ hyperfine line emission with a setup similar to GAS, \citet{Pineda_2010} observed a coherent core in the B5 region in Perseus and spatially resolved the ``transition to coherence,'' with NH$_3$ linewidths changing from supersonic values outside the core to subsonic values inside.  The coherent core is included in the analysis presented in Paper I.  However, the internal velocity structure of B5, derived from NH$_3$ line fitting, cannot be well described by a linear velocity gradient.  Thus, the coherent core in B5, while being a member of the dense cores traced by NH$_3$ emission, is not included in the following analyses.

\section{Analysis}
\label{sec:analysis}

\subsection{Velocity Gradient}
\label{sec:analysis_gradient}
\begin{figure}[ht!]
\epsscale{0.74}
\plotone{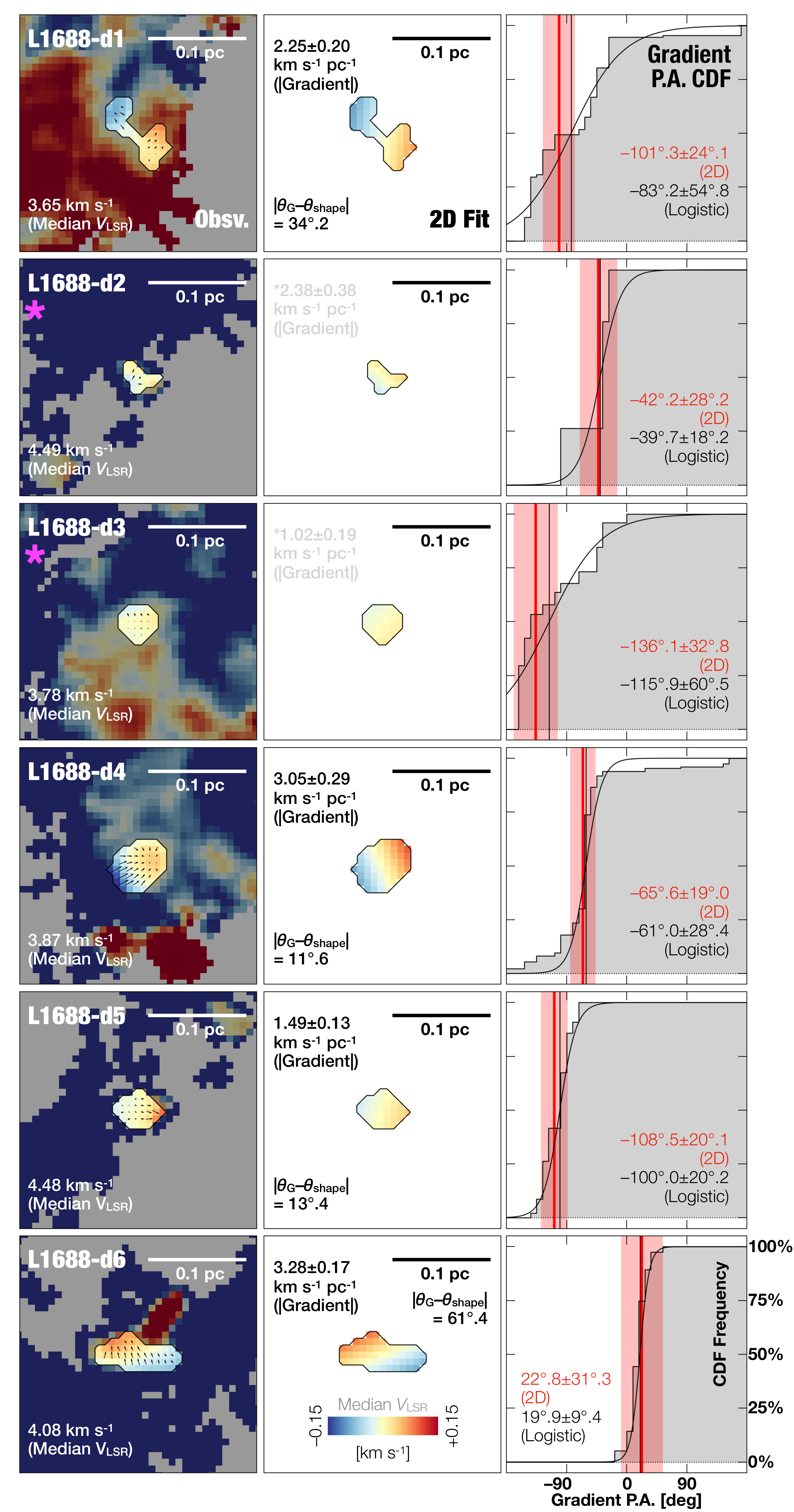}
\caption{\label{fig:gallery1} Observed NH$_3$ velocity centroids ($V_\mathrm{LSR}$), the 2D linear fit, and the cumulative distribution function (CDF) of pixel-by-pixel velocity gradient position angles, for L1688-d1 to L1688-d6.  L1688-d2 and L1688-d3, marked with magenta asterisks, are excluded from subsequent analyses in this paper, due to the small number of available pixels.  \textit{Left:}  Observed $V_\mathrm{LSR}$ of NH$_3$ emission.  The contour marks the droplet boundary.  The scale bar corresponds to 0.1 pc at the distance of Ophiuchus.  The color scale of each panel has the same stretch, ranging from $-$0.15 km s$^{-1}$ to +0.15 km s$^{-1}$ from the median $V_\mathrm{LSR}$ of each droplet.  \textit{Center:}  Fitted $V_\mathrm{LSR}$, based on the 2D linear fit (see \S\ref{sec:analysis_gradient}).  The color scale is the same as that used for the corresponding panel in the left column.  \textit{Right:}  Cumulative distribution function (CDF) of the pixel-by-pixel velocity gradient position angles, based on the change in $V_\mathrm{LSR}$ across neighboring pixels.  The red vertical line shows the position angle of the gradient from the 2D linear fit.  The black curve is a logistic function fitted to the CDF, and the vertical black line shows the midpoint of the fitted logistic function.}
\end{figure}

\begin{figure}[ht!]
\epsscale{0.74}
\plotone{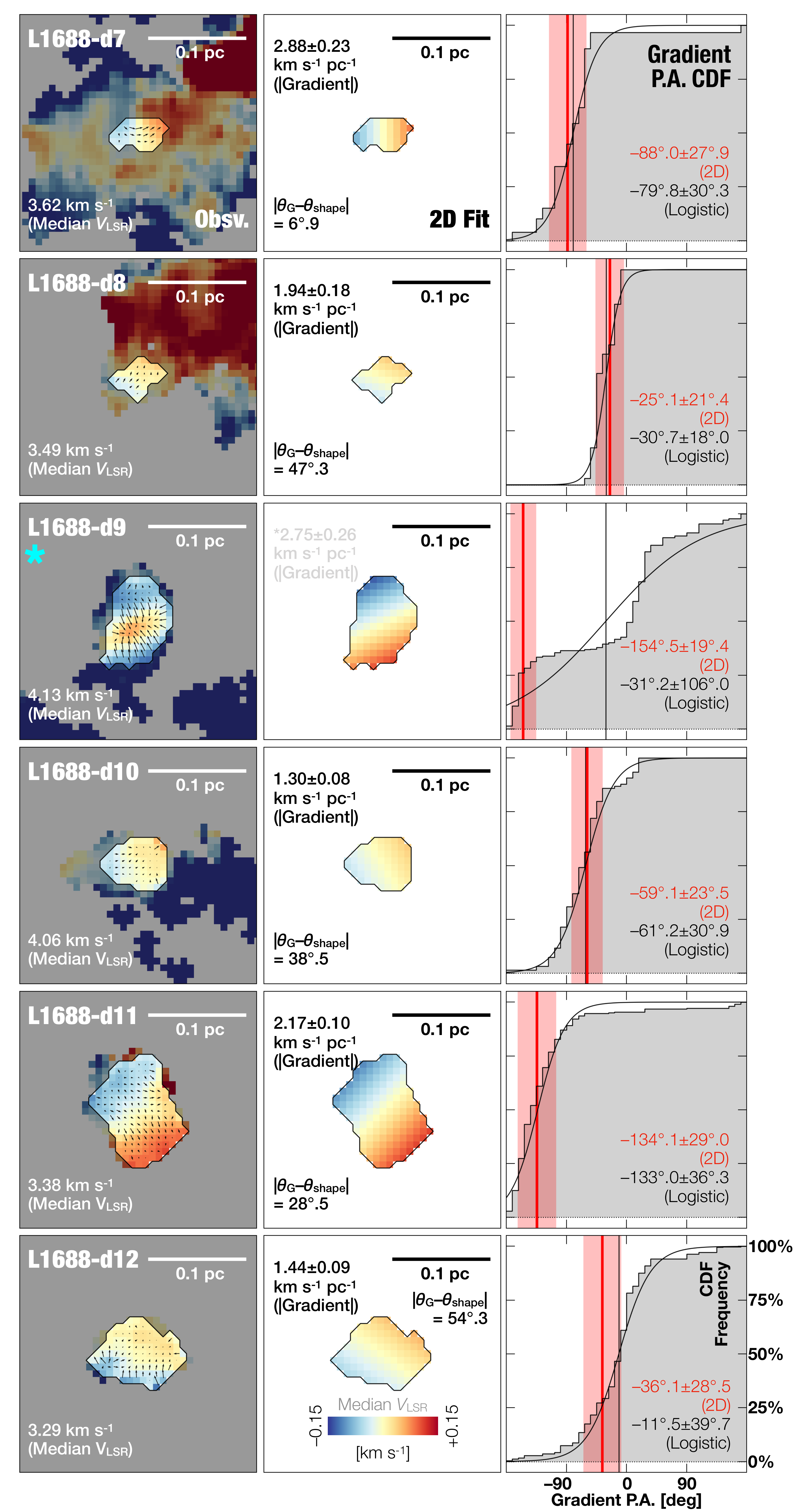}
\caption{\label{fig:gallery2} Same as Fig.\ \ref{fig:gallery1} for L1688-d7 to L1688-d12.  L1688-d9, marked with a cyan asterisk, is excluded from subsequent analyses in this paper, because of the non-linear distribution of $V_\mathrm{LSR}$ for which a linear velocity fit cannot be generated without a high uncertainty.  Notice that the non-linear distributions of $V_\mathrm{LSR}$ of L1688-d9 can be identified in their CDFs (in the right column).}
\end{figure}

\begin{figure}[ht!]
\epsscale{0.74}
\plotone{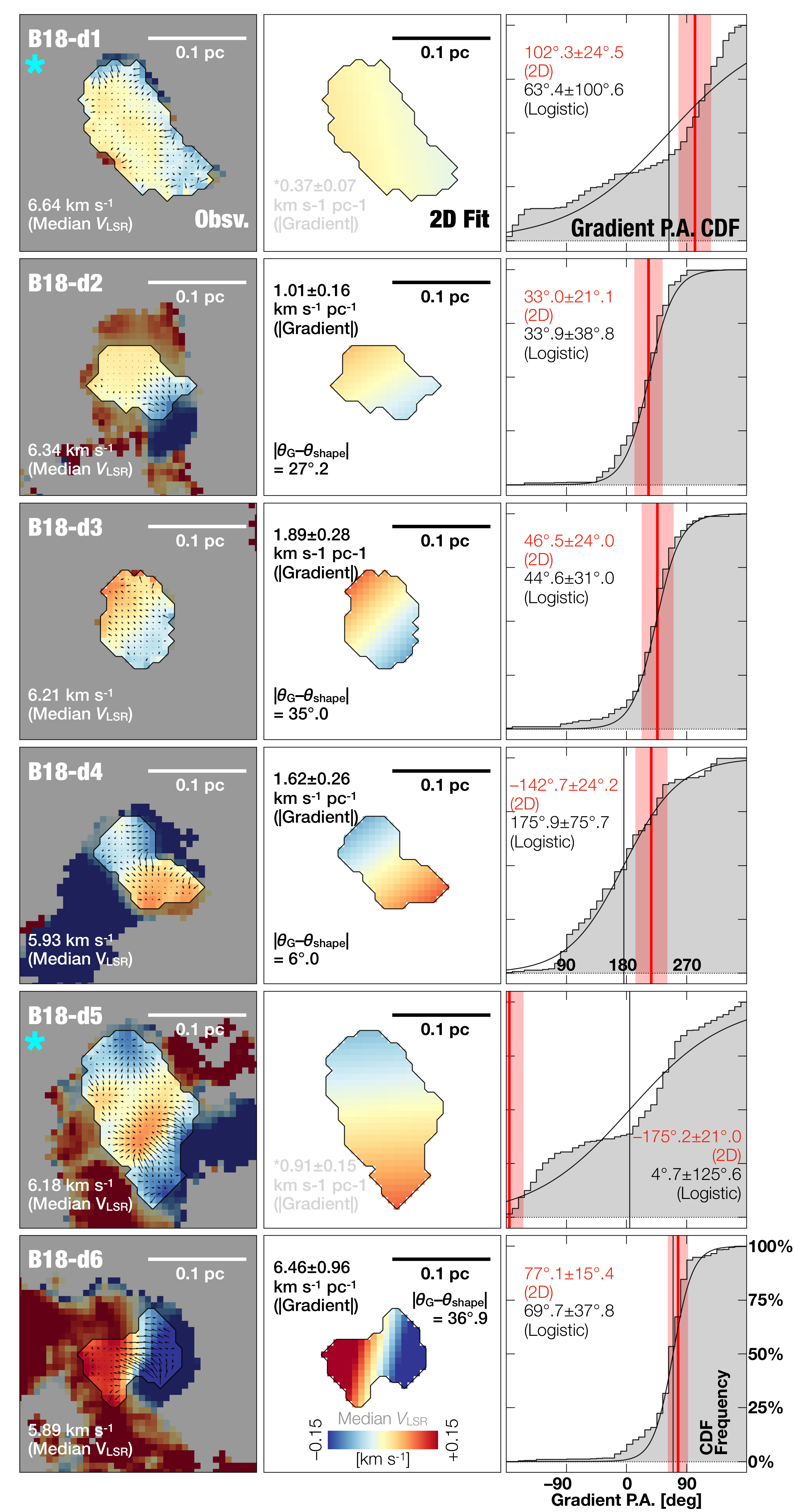}
\caption{\label{fig:gallery3} Same as Fig.\ \ref{fig:gallery1} for B18-d1 to B18-d6.  B18-d1 and B18-d5, marked with cyan asterisks, are excluded from subsequent analyses in this paper, because of the non-linear distributions of $V_\mathrm{LSR}$ for which a linear velocity fit cannot be generated without a high uncertainty.  Notice that the non-linear distributions of $V_\mathrm{LSR}$ of B18-d1 and B18-d5 can be identified in their CDFs (in the right column).  Since for B18-d4, the fitted velocity gradient and the distribution of local velocity gradients have typical position angles close to the -180/180 degrees extremes, the CDF is shown for a range from 0 to 360 degrees (East of North) instead.  The CDFs of other droplets are shown for a range from -180 to 180 degrees, as in Fig.\ \ref{fig:gallery1} and \ref{fig:gallery2}.}
\end{figure}

We adopt the method used by G93 and fit the observed velocity structure of each droplet with a 2D linear velocity distribution.  That is, the observed map of $V_\mathrm{LSR}$ (from NH$_3$ hyperfine line fitting; see \S\ref{sec:data_GAS_fitting}) is fit with a two-dimensional first-order linear function (a plane in the position-position-velocity space):

\begin{equation}
\label{eq:linfit}
v(x, y) = c_x x + c_y y + c_0\ \mathrm{,}
\end{equation}

\noindent where $x$ and $y$ stand for the position of each pixel in the plane of the sky on the $V_\mathrm{LSR}$ map in physical length units, and $c_x$, $c_y$, and $c_0$ are constant coefficeints.  The fit is carried out using an Astropy implementation of the Levenberg-Marquardt algorithm of a least squares regression analysis.  Then, the linear velocity gradient, $\mathcal{G}$, is a vector:

\begin{equation}
\label{eq:gradient}
\mathcal{G} = (c_x, c_y)\ \mathrm{,}
\end{equation}

\noindent which has a magnitude, $\left|\mathcal{G}\right| = \sqrt{c_x^2 + c_y^2}$, and an orientation, $\theta_\mathcal{G} = \arctan{(c_y/c_x)}$\footnote{In this paper, directions in the plane of the sky are expressed in position angles, with the North as the origin and increasing from the North to the East.}.  The uncertainties of the observed velocity centroid and the sampling on the pixel grid are propagated to the uncertainty in the measured gradient presented in Table \ref{table:gradient}.  Fig.\ \ref{fig:gallery1}, \ref{fig:gallery2}, and \ref{fig:gallery3} show the fitted linear velocity fields for the droplets in L1688 and B18 (middle-column panels), in comparison to the observed $V_\mathrm{LSR}$ (left-column panels).  Table \ref{table:gradient} lists the resultant velocity gradient magnitudes and orientations.

% 3-row plots.

%[width=1\textwidth]

\begin{longrotatetable}
\begin{deluxetable*}{lcccccc}
\tablecaption{Velocity Gradient and Rotational Motion of Droplets and Droplet Candidates\label{table:gradient}}
\tablehead{\colhead{ID\tablenotemark{a}} & \colhead{Grad.\ Magnitude\tablenotemark{b}} & \colhead{Grad.\ Orientation\tablenotemark{b}} & \colhead{Specific Ang.\ Momentum\tablenotemark{c}} & \colhead{Rotational Energy\tablenotemark{c}} & \colhead{} & \colhead{Shape Orientation\tablenotemark{e}} \\ \colhead{} & \colhead{($\left|\mathcal{G}\right|$)} & \colhead{($\theta_\mathcal{G}$)} & \colhead{($J/M$)} & \colhead{($E_\mathrm{rot}$)} & \colhead{$\beta$\tablenotemark{d}} & \colhead{($\theta_\mathrm{shape}$)} \\ \colhead{} & \colhead{km s$^{-1}$ pc$^{-1}$} & \colhead{deg.\ E of N} & \colhead{km s$^{-1}$ pc} & \colhead{erg} & \colhead{} & \colhead{deg.\ E of N}}
\startdata
L1688-d1 & $2.25\pm 0.20$ & $-101\pm 24$ & $1.3\pm 0.5\times 10^{-3}$ & $5.1\pm 2.3\times 10^{39}$ & $0.13\pm 0.08$ & $45$ \\ 
L1688-d4 & $3.05\pm 0.29$ & $-66\pm 19$ & $1.3\pm 0.6\times 10^{-3}$ & $2.8\pm 1.4\times 10^{40}$ & $0.03\pm 0.02$ & $-54$ \\
L1688-d5 & $1.49\pm 0.13$ & $-109\pm 20$ & $3.6\pm 2.1\times 10^{-4}$ & $6.9\pm 4.4\times 10^{38}$ & $0.02\pm 0.02$ & $85$ \\
L1688-d6 & $3.28\pm 0.17$ & $23\pm 31$ & $1.3\pm 0.6\times 10^{-3}$ & $9.8\pm 4.8\times 10^{39}$ & $0.12\pm 0.09$ & $84$ \\
L1688-d7 & $2.88\pm 0.23$ & $-88\pm 28$ & $8.0\pm 4.4\times 10^{-4}$ & $2.4\pm 1.4\times 10^{39}$ & $0.11\pm 0.10$ & $-81$ \\
L1688-d8 & $1.94\pm 0.18$ & $-25\pm 21$ & $5.2\pm 2.9\times 10^{-4}$ & $9.8\pm 5.9\times 10^{38}$ & $0.05\pm 0.05$ & $-72$ \\
L1688-d10 & $1.30\pm 0.08$ & $-59\pm 24$ & $5.9\pm 2.6\times 10^{-4}$ & $1.7\pm 0.8\times 10^{39}$ & $0.02\pm 0.02$ & $82$ \\
L1688-d11 & $2.17\pm 0.10$ & $-134\pm 29$ & $2.5\pm 0.7\times 10^{-3}$ & $2.5\pm 0.7\times 10^{40}$ & $0.12\pm 0.05$ & $17$ \\
L1688-d12 & $1.44\pm 0.09$ & $-36\pm 29$ & $1.0\pm 0.4\times 10^{-3}$ & $5.6\pm 2.1\times 10^{39}$ & $0.03\pm 0.02$ & $90$ \\
\hline
B18-d2 & $1.01\pm 0.16$ & $33\pm 21$ & $8.0\pm 2.7\times 10^{-4}$ & $1.1\pm 0.5\times 10^{40}$ & $0.01\pm 0.01$ & $60$ \\
B18-d3 & $1.89\pm 0.28$ & $47\pm 24$ & $1.5\pm 0.5\times 10^{-3}$ & $1.6\pm 0.7\times 10^{40}$ & $0.04\pm 0.02$ & $12$ \\
B18-d4 & $1.62\pm 0.26$ & $-143\pm 24$ & $1.6\pm 0.5\times 10^{-3}$ & $1.7\pm 0.7\times 10^{40}$ & $0.04\pm 0.02$ & $43$ \\
B18-d6 & $6.46\pm 0.96$ & $77\pm 15$ & $6.0\pm 1.9\times 10^{-3}$ & $3.2\pm 1.3\times 10^{41}$ & $0.44\pm 0.23$ & $-66$ \\
\hline
L1688-c1E\tablenotemark{f} & $2.32\pm 0.31$ & $-131\pm 27$ & $3.7\pm 2.7\times 10^{-4}$ & $1.4\pm 1.4\times 10^{38}$ & $0.21\pm 0.21$ & \nodata \\
L1688-c1W\tablenotemark{f} & $1.33\pm 0.28$ & $-59\pm 29$ & $2.5\pm 1.7\times 10^{-4}$ & $2.2\pm 1.9\times 10^{38}$ & $0.02\pm 0.02$ & \nodata \\
L1688-c2\tablenotemark{f} & $1.44\pm 0.40$ & $113\pm 30$ & $3.4\pm 2.2\times 10^{-4}$ & $4.8\pm 4.1\times 10^{38}$ & $0.02\pm 0.02$ & \nodata \\
L1688-c3\tablenotemark{f} & $3.34\pm 0.30$ & $50\pm 14$ & $4.6\pm 3.6\times 10^{-4}$ & $9.9\pm 8.6\times 10^{38}$ & $0.09\pm 0.09$ & \nodata \\
L1688-c4\tablenotemark{f} & $1.70\pm 0.20$ & $-124\pm 20$ & $5.4\pm 2.8\times 10^{-4}$ & $4.5\pm 3.0\times 10^{38}$ & $0.10\pm 0.09$ & \nodata
\enddata
\tablenotetext{a}{L1688-d2 and L1688-d3 are excluded because of the small number of Nyquist-sampled pixels available to the linear velocity fit.  L1688-d9, B18-d1, and B18-d5 are excluded because of non-linear velocity structures, which result in linear fits with high uncertainties.}
\tablenotetext{b}{Estimated from 2D linear velocity fits and defined between $-180$ and $180$ degrees.  See \S\ref{sec:analysis_gradient}.}
\tablenotetext{c}{Estimated based on the fitted velocity gradient, assuming that a rotational motion leads to the velocity gradient.  The rotational motion is assumed to be solid-body, and the rotating body is assumed to have a uniform density.}
\tablenotetext{d}{The ratio between the rotational energy, $E_\mathrm{rot}$, and the absolute value of gravitational potential energy, $\left|\Omega_\mathrm{G}\right|$.  See \S\ref{sec:analysis_beta} for details.}
\tablenotetext{e}{The position angle of the major axis, derived from NH$_3$ brightness in a principal component analysis (PCA), and defined between $-90$ and $90$ degrees.  See Paper I for details.}
\tablenotetext{f}{L1688-c1E, L1688-c1W, L1688-c2, L1688-c3, and L1688-c4 are droplet candidates (see \S\ref{sec:data_catalogs_droplets}).  All droplet candidates satisfy the validation of the gradient fits described in \S\ref{sec:analysis_gradient}.  However, since they do not satisfy at least one of the criteria used to define droplets in Paper I, we do not include droplet candidates in the quantitative analysis presented in this paper.  They are included in the figures for reference.}
\end{deluxetable*}
\end{longrotatetable}

%\hline
%L1688-c1E\tablenotemark{f} & $2.32\pm 0.31$ & $-131$ & $3.7\pm 0.5\times 10^{-4}$ & $1.4\pm 1.7\times 10^{38}$ & $2.1\pm 2.7 \times 10^{-1}$ & $-13$ \\
%L1688-c1W\tablenotemark{g} & $1.33\pm 0.28$ & $-58$ & $2.5\pm 0.6\times 10^{-4}$ & $2.2\pm 1.1\times 10^{38}$ & $2.1\pm 1.1 \times 10^{-2}$ & $-16$ \\
%L1688-c2 & $1.44\pm 0.41$ & $112$ & $3.4\pm 1.0\times 10^{-4}$ & $4.8\pm 2.9\times 10^{38}$ & $2.3\pm 1.4 \times 10^{-2}$ & $-37$ \\
%L1688-c3 & $3.34\pm 0.30$ & $50$ & $4.6\pm 0.5\times 10^{-4}$ & $9.9\pm 4.1\times 10^{38}$ & $8.6\pm 3.6 \times 10^{-2}$ & $-14$ \\
%L1688-c4 & $1.70\pm 0.20$ & $-124$ & $5.4\pm 0.7\times 10^{-4}$ & $4.5\pm 2.0\times 10^{38}$ & $1.0\pm 0.5 \times 10^{-1}$ & $9$ \\
%\tablenotetext{f}{The eastern part of L1688-d1.}
%\tablenotetext{g}{The western part of L1688-d1.}

We validate our results from the linear velocity fit with the distributions of the local velocity gradients, measured from velocity change across neighboring pixels (arrows in the left-column panels of Fig.\ \ref{fig:gallery1}, \ref{fig:gallery2}, and \ref{fig:gallery3}).  In an ideal case where the observed $V_\mathrm{LSR}$ map can be fully described by a linear velocity field (Equation \ref{eq:linfit}), the pixel-by-pixel cumulative distribution function (CDF) for the local velocity gradient orientation would be a step function with the change (from 0 to 1) occurring at the orientation of the velocity gradient corresponding to the linear velocity field, $\theta_\mathcal{G}$ (Equation \ref{eq:gradient}).  Thus, the \textit{goodness} of the linear velocity fit can be estimated by examining the CDFs of local velocity gradient directions (see the right-column panels of Fig.\ \ref{fig:gallery1}, \ref{fig:gallery2}, and \ref{fig:gallery3}).  In the following analyses, we exclude the droplets where both the observed $V_\mathrm{LSR}$ and the CDFs show clear signs of non-linear velocity structures (L1688-d9, B18-d1, and B18-d5).  We also exclude L1688-d2 and L1688-d3 because of the small number of Nyquist-sampled pixels available to the linear velocity fit.

\begin{figure}[ht!]
\plotone{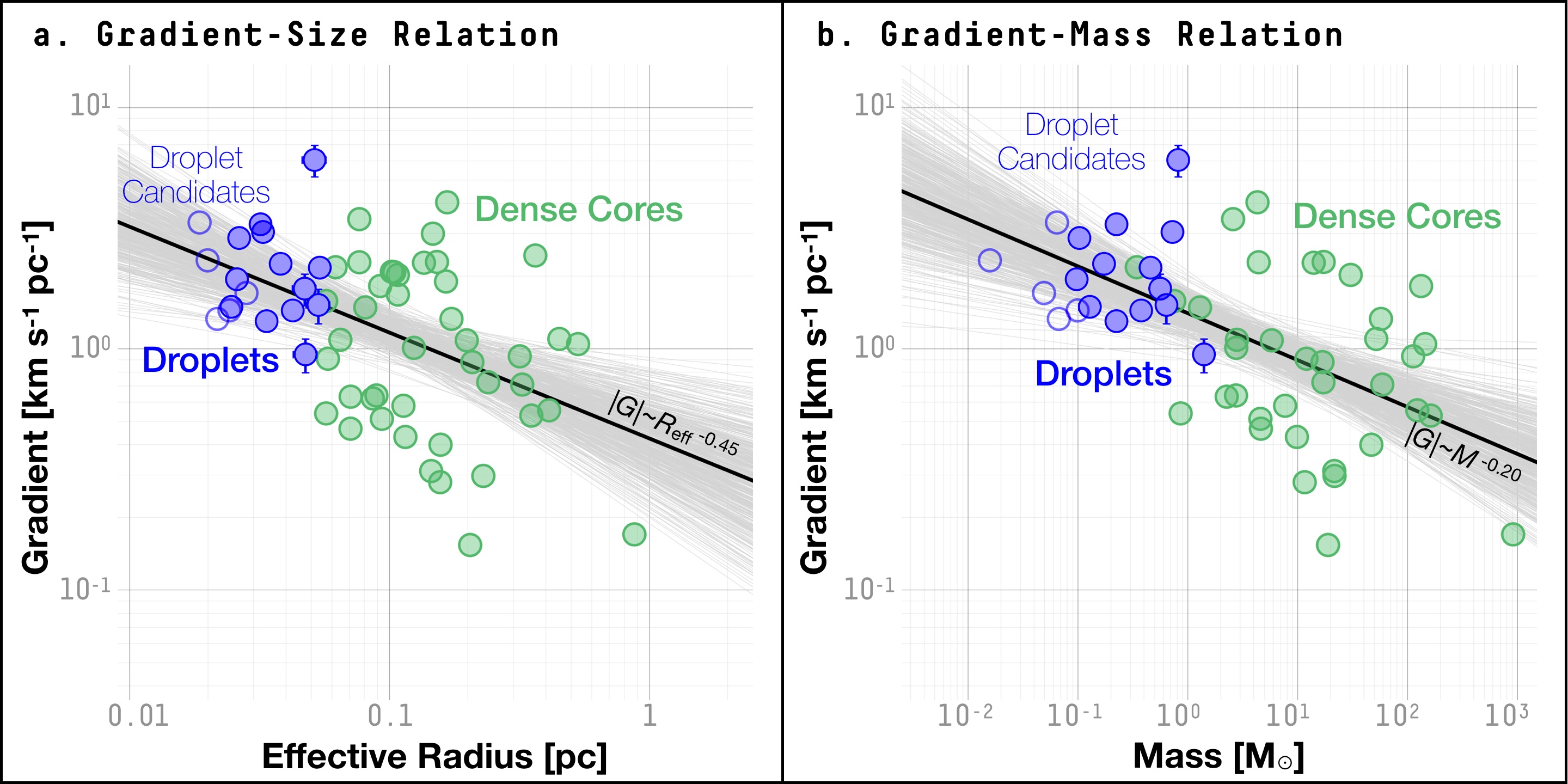}
\caption{\label{fig:Gradient} \textbf{(a)} Velocity gradient, plotted against the effective radius, for larger-scale dense cores (green dots), droplets (blue filled dots), and droplet candidates (blue empty dots).  The black line shows a power-law relation between the velocity gradient and the effective radius, found for all cores shown here (both dense cores and droplets; excluding droplet candidates) by a gradient-based MCMC sampler.  Randomly selected 10\% of the accepted parameters in the MCMC chain are plotted as gray lines as a reference of uncertainty in the fitting.  \textbf{(b)} Velocity gradient, plotted against the mass, for larger-scale dense cores (green dots), droplets (blue filled dots), and droplet candidates (blue empty dots).  Similar to (a), the black line shows a power-law relation between the velocity gradient and the effective radius, and randomly selected 10\% of the accepted parameters in the MCMC chain used to find the best-fit power law are plotted as gray lines as a reference of uncertainty in the fitting.}
\end{figure}

We find a typical value of velocity gradient magnitude of $1.94_{-0.51}^{+1.12}$ km s$^{-1}$ pc$^{-1}$ for droplets\footnote{Unless otherwise noted, the typical value of each physical property presented in this work is the median value of all the droplets with reliable outcomes of the linear velocity fitting---excluding the droplet candidates---with the upper and lower bounds being the values measured at the 84th and 16th percentiles, which would correspond to $\pm$1 standard deviation around the median value if the distribution is Gaussian.}.  Fig.\ \ref{fig:Gradient} shows the relations between the velocity gradient magnitude and the effective radius and mass (derived in Paper I).  Consistent with what G93 found for larger-scale dense cores, the velocity gradient magnitudes found for droplets appear to increase toward smaller size scales, although with a large dispersion in velocity gradient magnitude (Fig.\ \ref{fig:Gradient}a; note that the velocity gradient and the effective radius are mutually independently measured).  The droplets and larger-scale dense cores together appear to loosely follow a power-law relation between the velocity gradient magnitude and the size, $\left|\mathcal{G}\right| \propto R_\mathrm{eff}^{-0.45\pm 0.13}$.  G93 found $\left|\mathcal{G}\right| \propto R_\mathrm{eff}^{-0.4}$ for the larger-scale dense cores.  Since the dense cores and droplets follow a relatively tight mass-size relation, $M \propto R_\mathrm{eff}^{2.4\pm 0.1}$ (see Fig.\ 10 in Paper I), they \textit{consequently} appear to follow a power-law relation between the velocity gradient magnitude and the mass (see Fig.\ \ref{fig:Gradient}b).

% gradient-size/mass

%%%%%
Before converting the velocity gradient measured from the planar fit defined by Equation \ref{eq:linfit} to rotational properties, we would like to emphasize that the observed velocity patterns do not necessarily arise purely from rotational motions.  The rotational properties analyzed below in \S\ref{sec:analysis_JoverM} and \S\ref{sec:analysis_beta} should be treated as measurements of the ``net rotational motions'' that are the results of rotational motions, as well as (but not limited to) turbulence, gravitational infall, and larger-scale material flow.  That is, in an ideal situation where the observed $V_\mathrm{LSR}$ distribution is a perfect representation of the 3D motions, these rotational properties would capture any material movement that has a non-zero tangential component in a cylindrincal coordinate system centered at the core center of mass.  In reality, observational effects may also contribute to the measurements presented in the following analyses, such as the uncertainty in the inclination angle and the beam effect across the boundary of a core embedded in the molecular cloud (e.g.\ L1688-d4 shown in Fig.\ \ref{fig:gallery1} and L1688-d7 shown in Fig.\ \ref{fig:gallery2}).
%%%%%

\subsection{Specific Angular Momentum}
\label{sec:analysis_JoverM}
Assuming that the observed velocity gradient represents net rotational motion of the droplet, we can derive a specific angular momentum based on the velocity gradient.  Under this assumption, the angular velocity of the rotational motion is a function of the velocity gradient and inclination:

\begin{equation}
\label{eq:angvel}
\omega = \frac{\left|\mathcal{G}\right|}{\sin{i}}\ \mathrm{,}
\end{equation}

\noindent where $i$ is the inclination angle.  Following G93, we adopt $\sin{i} = 1$ in the following analyses, and thus the angular velocity has the same magnitude as the velocity gradient, $\omega = \left|\mathcal{G}\right|$.  This represents a lower limit on the angular velocity.  We can then estimate the rotational properties based on the fitted velocity gradient.

\begin{figure}[ht!]
\plotone{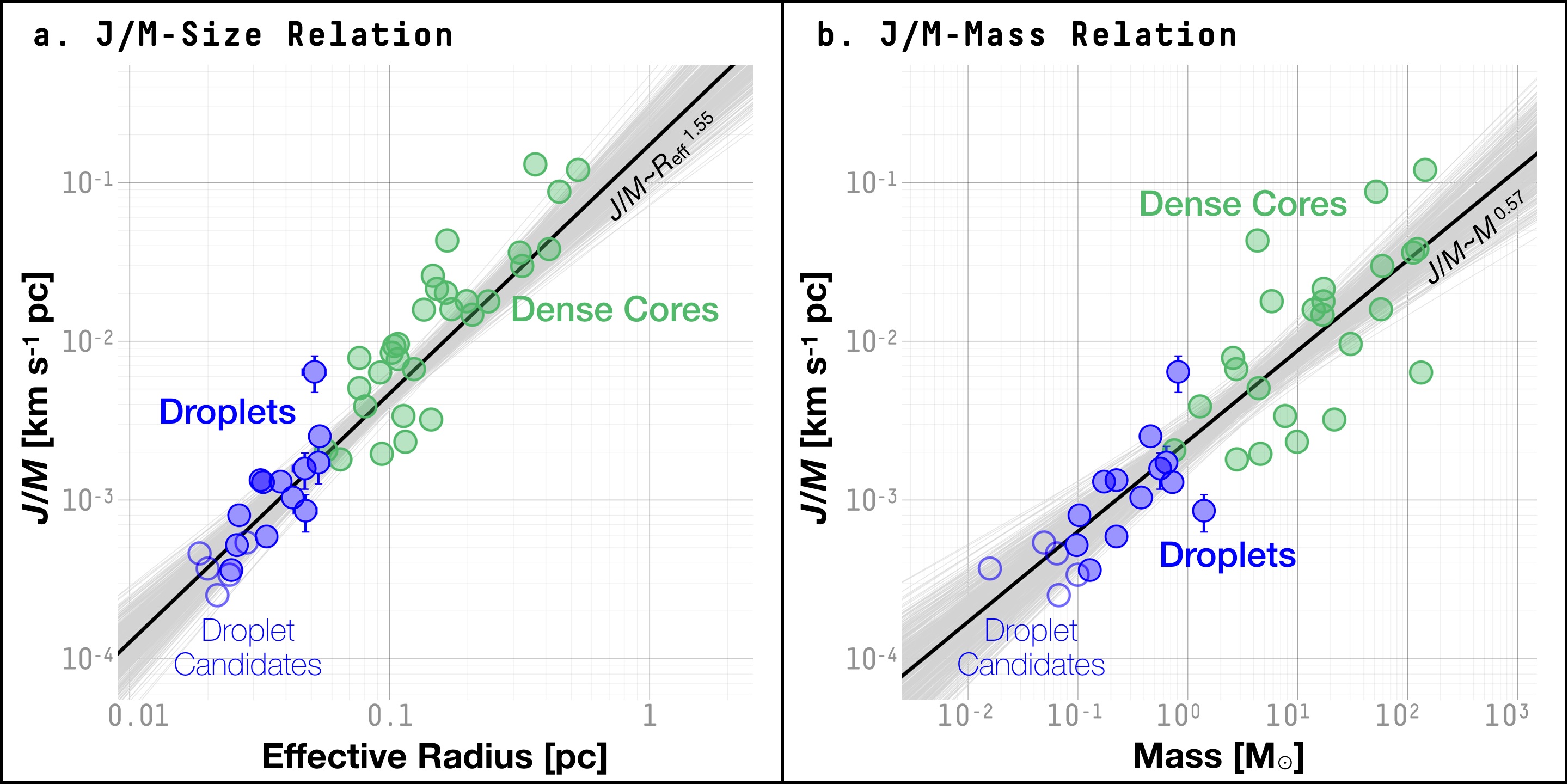}
\caption{\label{fig:JoverM} \textbf{(a)} Specific angular momentum, $J/M$, plotted against the effective radius, for larger-scale dense cores (green dots), droplets (blue filled dots), and droplet candidates (blue empty dots).  The black line shows a power-law relation between the velocity gradient and the effective radius, found for all cores shown here (both dense cores and droplets; excluding droplet candidates) by a gradient-based MCMC sampler.  Randomly selected 10\% of the accepted parameters in the MCMC chain are plotted as gray lines as a reference of uncertainty in the fitting.  \textbf{(b)} Specific angular momentum, $J/M$, plotted against the mass, for larger-scale dense cores (green dots), droplets (blue filled dots), and droplet candidates (blue empty dots).  Similar to (a), the black line shows a power-law relation between the velocity gradient and the effective radius, and randomly selected 10\% of the accepted parameters in the MCMC chain used to find the best-fit power law are plotted as gray lines as a reference of uncertainty in the fitting.}
\end{figure}

% moment of inertia
If the rotational motion giving rise to the observed velocity gradient is represented by solid-body rotation and the rotating body has a spherical geometry with a uniform density, the moment of inertia around its rotational axis is

\begin{equation}
\label{eq:I}
I = \frac{2}{5} MR^2\ \mathrm{,}
\end{equation}

\noindent where $M$ and $R$ are the mass and radius of the rotating object, respectively.  While assuming that all cores are spherical and have uniform densities can be unrealistic, it has been shown in Paper I that the droplets have small aspect ratios ($\lesssim$ 2; projected on the plane of the sky) and nearly uniform densities within their boundaries.  In this work, we adopt the effective radius of each droplet/dense core for $R$, derived from the geometric mean of the major and minor axes based on a principal component analysis of the NH$_3$ brightness distribution (see G93 and Paper I).  We would like to note that most droplets have aspect ratios between 1 and 2, with the exceptions of L1688-d1 and L1688-d6, both of which have aspect ratios close to 2.5.  Since angular momentum $J = I \omega$, the specific angular momentum, defined as the ratio between the angular momentum and the mass, is

\begin{equation}
\label{eq:JoverM}
\frac{J}{M} = \frac{2}{5} \omega R^2\ \mathrm{,}
\end{equation}

\noindent where $\omega$ is the angular velocity, which can be estimated from the observed velocity gradient (Equation \ref{eq:angvel}).  Using Equation \ref{eq:JoverM}, we find a typical value of $J/M = 1.3_{-0.7}^{+0.5}\times 10^{-3}$ km s$^{-1}$ pc for the droplets.  Fig.\ \ref{fig:JoverM} shows the relations between the specific angular momentum, $J/M$, and the size and mass of each droplet/dense core.  For both the droplets and the dense cores, we find

\begin{equation}
\label{eq:JoverM_powerlaw}
\frac{J}{M} = 10^{-0.72\pm 0.20} \left(\frac{R}{\mathrm{1 pc}}\right)^{1.55\pm 0.20}\ \mathrm{km}\ \mathrm{s^{-1}}\ \mathrm{pc,}
\end{equation}

\noindent and the relation between the specific angular momentum and the size derived by G93 for the dense cores extends to smaller sizes.  However, as G93 has noticed, the seemingly tight power-law relation between $J/M$ and the radius is partly due to the fact that $J/M$ is a power-law function of the radius (Equation \ref{eq:JoverM}).  And, since the droplets and dense cores follow a power-law relation between the size and the mass, they appear to follow a power-law relation between $J/M$ and the mass.

% J/M-mass/size

Note that in this paper, we follow G93 and assume that the rotational motion giving rise to the observed velocity gradient is a solid-body rotation, and that the rotating body has a spherical geometry with a uniform density.  However, we do not suggest that the rotational motion is necessarily solid-body nor that the density structure is uniform or spherical.  The results presented in this paper can be compared to physical properties calculated under assumptions of rotational and/or density profiles other than the ones used in this work by applying the corrections derived by \citet{Pineda_2019} \citep[see Appendix C in][]{Pineda_2019}.

\subsection{Rotational Energy}
\label{sec:analysis_beta}
For a rotating body with a mass, $M$, and a radius, $R$, we can also estimate the rotational energy, $E_\mathrm{rot}$, of the rotational motion giving rise to the observed velocity gradient:

\begin{equation}
\label{eq:Erot}
E_\mathrm{rot} = \frac{1}{2}I \omega^2 = \frac{1}{5} M R^2 \omega^2\ \mathrm{,}
\end{equation}

\noindent assuming solid-body rotation and uniform density.  \citep[See Appendix C in][for detailed derivation of a general expression for $E_\mathrm{rot}$.]{Pineda_2019}  We find a typical value of $E_\mathrm{rot} = 10^{40^{+0.3}_{-0.8}}$ erg for the droplets.
% the range of $E_\mathrm{rot}$ is $9.8\pm 8.1\times 10^{39}$ erg.

\begin{figure}[ht!]
\plotone{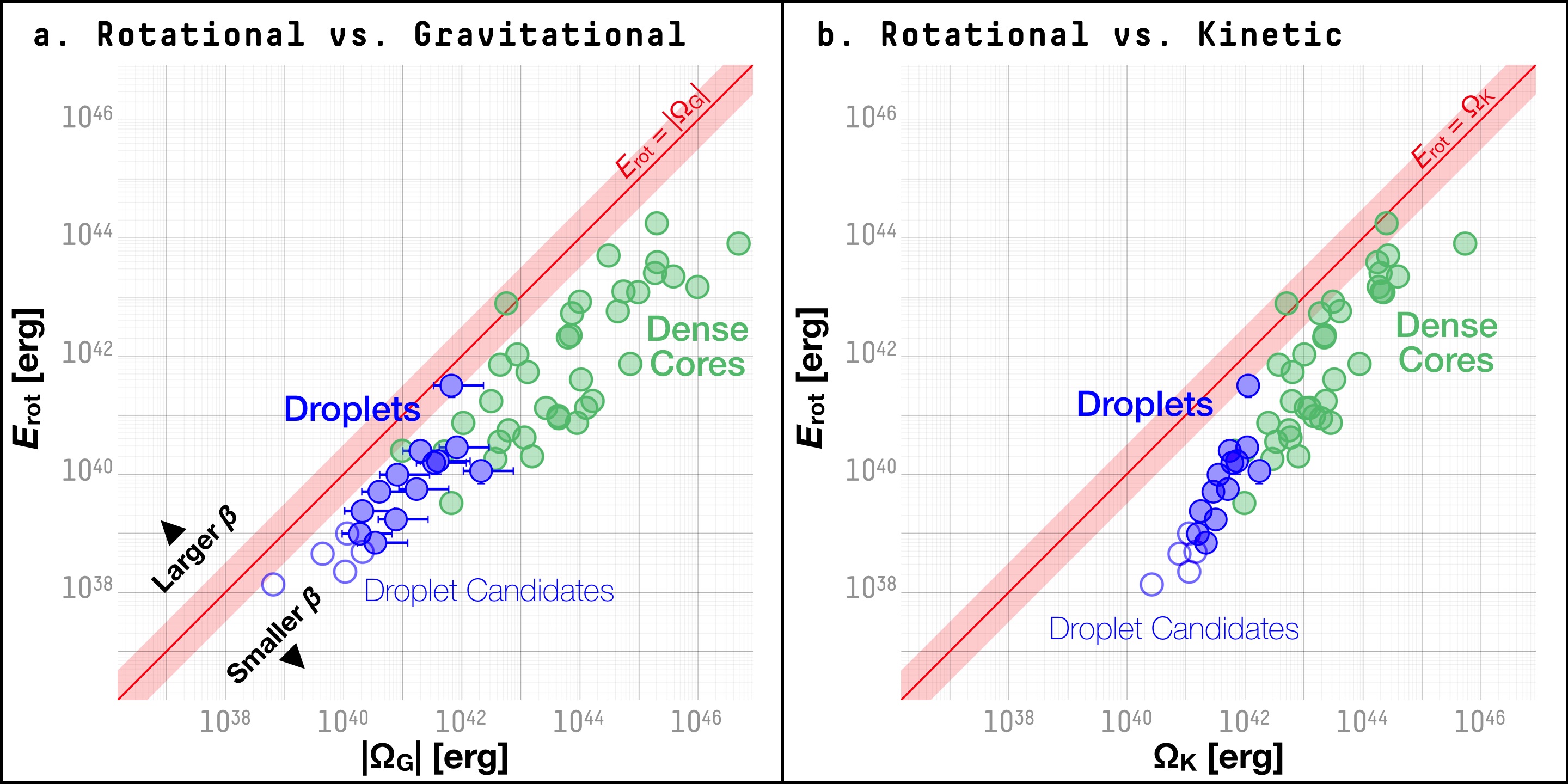}
\caption{\label{fig:Erot} \textbf{(a)} Rotational energy, $E_\mathrm{rot}$, plotted against the gravitational potential energy, $\left|\Omega_\mathrm{G}\right|$, for larger-scale dense cores (green dots), droplets (blue filled dots), and droplet candidates (blue empty dots).  The red line corresponds to the relation, $E_\mathrm{rot} = \left|\Omega_\mathrm{G}\right|$, and the red band marks the parameter space within an order of magnitude from this relation.  By definition, the ratio between the rotational and gravitational energies, $\beta$, is larger toward the top-left of the figure and smaller toward the bottom-right of the figure.  \textbf{(b)} Rotational energy, $E_\mathrm{rot}$, plotted against the kinetic energy, $\Omega_\mathrm{K}$, for larger-scale dense cores (green dots), droplets (blue filled dots), and droplet candidates (blue empty dots).  The red line corresponds to the relation, $E_\mathrm{rot} = \Omega_\mathrm{K}$, and the red band marks the parameter space within an order of magnitude from this relation.}
\end{figure}

In a fashion similar to a virial analysis, the rotational energy can be compared to the gravitational potential energy, $\Omega_\mathrm{G}$, of the rotating body\footnote{In this work, we adopt the same notions used in Paper I.}:

\begin{equation}
\label{eq:Gterm}
\Omega_\mathrm{G} = -\frac{3}{5}\frac{GM^2}{R}\ \mathrm{,}
\end{equation}

\noindent again assuming that the rotating body has a uniform density.  In comparison, a sphere of material with a power-law density distribution, $\rho \propto r^{-2}$, has an absolute value of gravitational potential energy, $\left|\Omega_\mathrm{G}\right|$, a factor of $\sim$ 1.7 larger than that expressed in Equation \ref{eq:Gterm}, and a sphere with a Gaussian density distribution has $\left|\Omega_\mathrm{G}\right|$ a factor of $\sim$ 2 smaller than that expressed in Equation \ref{eq:Gterm} \citep{Pattle_2015, Kirk_2017b}.  Similar to Paper I, in the following analysis, we include the deviation in $\Omega_\mathrm{G}$ due to different assumptions of density distributions in the estimated errors.  Fig.\ \ref{fig:Erot}a shows that the rotational energy is generally smaller than the absolute value of the gravitational potential energy by a order of magnitude or more, which suggests that self-gravity alone can provide the needed binding to sustain the rotational motion.  Fig.\ \ref{fig:Erot}a also suggests that the ratio between the rotational energy and the gravitational potential energy remains roughly constant.

% Erot-Gterm/Kterm

The ratio between the rotational energy and the gravitational potential energy, $\beta \equiv E_\mathrm{rot}/\Omega_\mathrm{G}$, is sometimes referred to as the ``rotational parameter'' \citep[e.g.][]{Dib_2010}.  The ratio between the rotational and gravitational energies is often taken as an input parameter to set up the initial conditions for disk and planet formation models \citep[e.g.][]{Allen_2003, Li_2011, Seifried_2011, Seifried_2012}, especially to scale the angular velocity of the rotation with respect to the size and mass of the model.  From Equation \ref{eq:Erot} and \ref{eq:Gterm}, we can derive $\beta$:

\begin{equation}
\label{eq:beta}
\beta \equiv \frac{E_\mathrm{rot}}{\Omega_\mathrm{G}} = \frac{1}{3} \frac{\omega^2 R^3}{GM}\ \mathrm{,}
\end{equation}

\noindent with the same assumptions of solid-body rotation and a uniform density, where $\omega$ is the angular velocity (see Equation \ref{eq:angvel}).  For the droplets, we find a typical value of $\beta = 0.046_{-0.024}^{+0.079}$, compared to $\beta \sim 0.032$ found by G93 for larger-scale dense cores.  Including both the droplets and larger-scale dense cores analyzed by G93, we find $\beta \sim 0.039$.
%%% S4.3

\begin{figure}[ht!]
\plotone{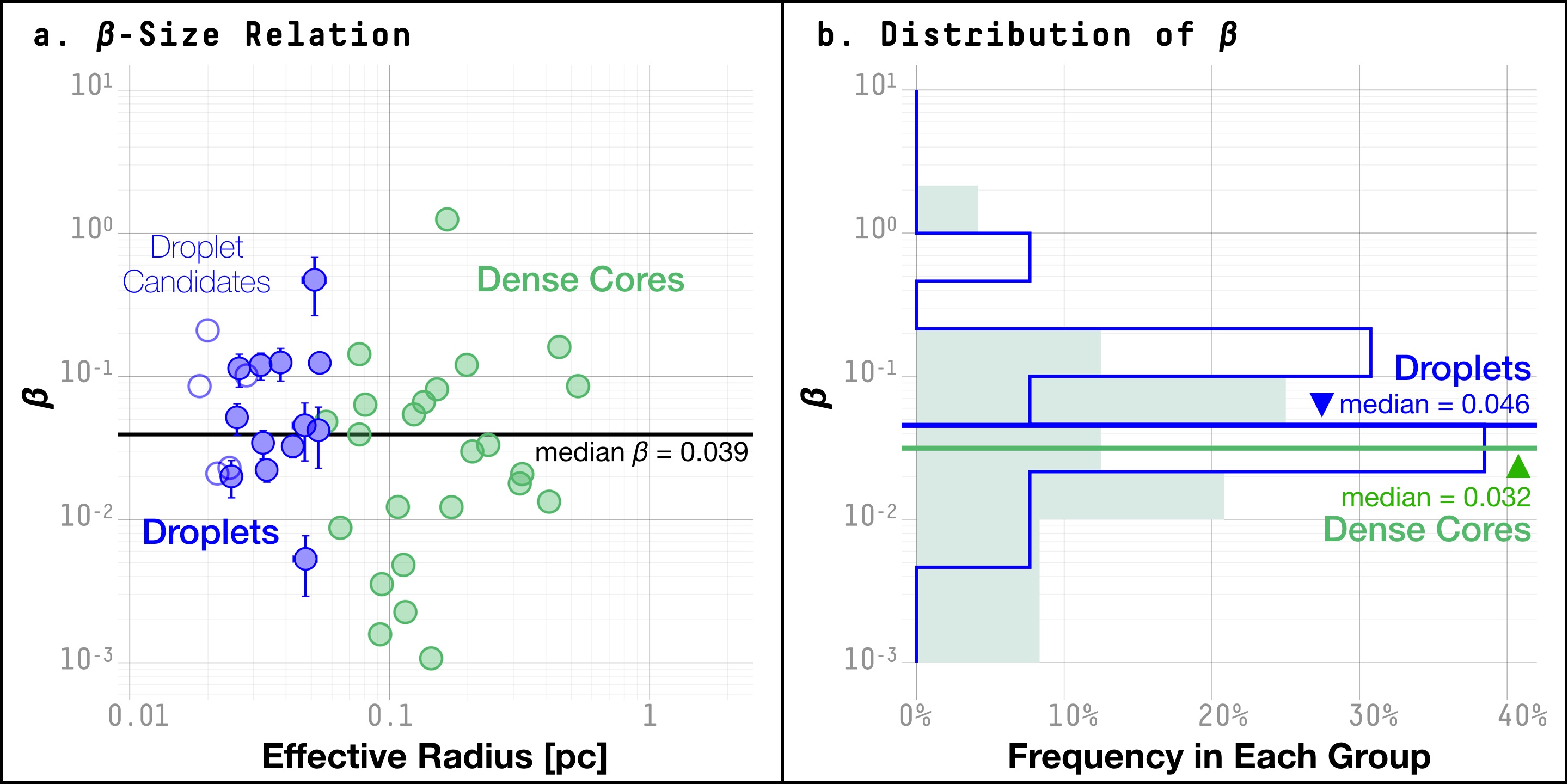}
\caption{\label{fig:beta} \textbf{(a)} Ratio between rotational and gravitational energies, $\beta$, plotted against the effective radius, for larger-scale dense cores (green dots), droplets (blue filled dots), and droplet candidates (blue empty dots).  The horizontal black line marks the median value of $\beta$ for all cores shown in the figure (both dense cores and droplets; excluding droplet candidates).  \textbf{(b)} Distribution of the ratio between rotational and gravitational energies, $\beta$, for larger-scale dense cores (green filled histogram) and droplets (excluding droplet candidates; blue histogram).  The solid horizontal lines mark the median value of each group.}
\end{figure}

Fig.\ \ref{fig:beta}a shows the relation between $\beta$ and the size of each droplet/dense core.  Consistent with what G93 found for larger-scale dense cores with observations of NH$_3$ emission, $\beta$ measured for the droplets appear to be independent of the size scale (Fig.\ \ref{fig:beta}a).  The distributions of $\beta$ for droplets and dense cores are statistically consistent with each other, despite difference in physical properties between the two groups of objects (Fig.\ \ref{fig:beta}b).  As G93 point out, this might be the result of the fact that the NH$_3$ emission traces almost constant-density gas.

% beta-size and distribution

%%%%%

Similarly, the rotational energy can be compared to the kinetic energy arising from the internal gas motions:

\begin{equation}
\label{eq:Kterm}
\Omega_\mathrm{K} = \frac{3}{2}M \sigma_\mathrm{tot}^2\ \mathrm{,}
\end{equation}

\noindent where $\sigma_\mathrm{tot}$ includes both the thermal and the turbulent motions of the gas inside a droplet or a larger-scale dense core:

\begin{equation}
\label{eq:SigmaTot}
\sigma_\mathrm{tot}^2 = \sigma_\mathrm{NT}^2 + \frac{k_\mathrm{B} T_\mathrm{kin}}{m_\mathrm{ave}}\ \mathrm{,}
\end{equation}

\noindent where $\sigma_\mathrm{NT}$ is the non-thermal (turbulent) velocity dispersion; $T_\mathrm{kin}$ is the kinetic energy; and $m_\mathrm{ave}$ is the average particle mass (2.37 u; \citealt{Kauffmann_2008}; Paper I).  Fig.\ \ref{fig:Erot}b shows that the rotational energy is generally smaller than the kinetic energy arising from the gas motions inside the cores.

%The result suggests that the rotational velocity at the outer edge of a core, $V_\mathrm{rot} = \omega R$ (Equation \ref{eq:Erot}), is smaller than the total velocity dispersion of the gas within the core, $\sigma_\mathrm{tot}$ (Equation \ref{eq:SigmaTot}).  Contrary to the comparison between the rotational energy and the gravitational potential energy, Fig.\ \ref{fig:Erot}b appears to suggest a change in the ratio between the rotational energy and the total kinetic energy from droplets to larger-scale cores.

%In general, the ratio between the rotational energy and the total kinetic energy, $E_\mathrm{rot}/\Omega_\mathrm{K}$, appears to be smaller for droplets than for the dense cores.  Fig.\ \ref{fig:ratioRotKin}a shows that, while there is a large scatter in the distribution of $E_\mathrm{rot}/\Omega_\mathrm{K}$ similar to $\beta$, a relation between size and $E_\mathrm{rot}/\Omega_\mathrm{K}$ is statistically more significant than the one between size and $\beta$.  The larger, gravitationally bound dense cores generally have larger $E_\mathrm{rot}/\Omega_\mathrm{K}$, and the smaller, pressure-bound droplets have smaller $E_\mathrm{rot}/\Omega_\mathrm{K}$.  Fig.\ \ref{fig:ratioRotKin} shows the overall distributions of $E_\mathrm{rot}/\Omega_\mathrm{K}$ for the two groups of objects.  The median value of $E_\mathrm{rot}/\Omega_\mathrm{K}$ of the dense cores is twice as large as the median value of $E_\mathrm{rot}/\Omega_\mathrm{K}$ of the droplets.

\begin{figure}[ht!]
\plotone{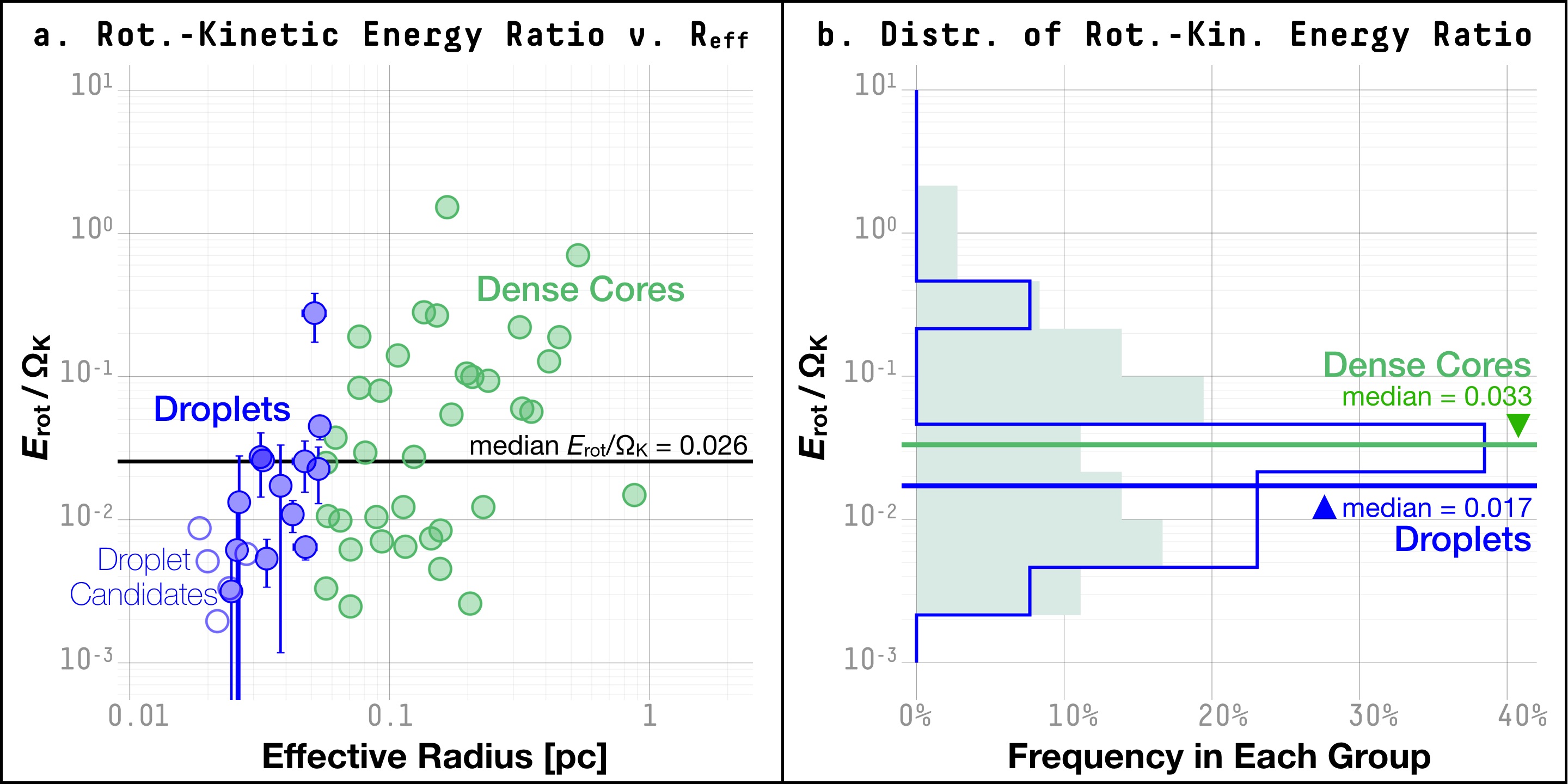}
\caption{\label{fig:ratioRotKin} \textbf{(a)} Ratio between rotational and total kinetic energies, $E_\mathrm{rot}/\Omega_\mathrm{K}$, plotted against the effective radius, for larger-scale dense cores (green dots), droplets (blue filled dots), and droplet candidates (blue empty dots).  The horizontal black line marks the median value of $E_\mathrm{rot}/\Omega_\mathrm{K}$ for all cores shown in the figure (both dense cores and droplets; excluding droplet candidates).  \textbf{(b)} Distribution of the ratio between rotational and gravitational energies, $E_\mathrm{rot}/\Omega_\mathrm{K}$, for larger-scale dense cores (green filled histogram) and droplets (excluding droplet candidates; blue histogram).  The solid horizontal lines mark the median value of each group.}
\end{figure}

Fig.\ \ref{fig:ratioRotKin}a shows that, similar to $\beta$, there is a large scatter in the distribution of $E_\mathrm{rot}/\Omega_\mathrm{K}$.  Relatively speaking, a relation between the size and $E_\mathrm{rot}/\Omega_\mathrm{K}$ is statistically more significant than the one between the size and $\beta$, indicated by a correlation index of 0.87 (versus 0.02 for the size-$\beta$ distribution) in a Pearson correlation test.  However, we note that a Pearson correlation coefficient of 0.87 cannot be used to determine the existence of a correlation by itself.  Fig.\ \ref{fig:ratioRotKin}b shows the overall distributions of $E_\mathrm{rot}/\Omega_\mathrm{K}$ for the two groups of objects.  The median value of $E_\mathrm{rot}/\Omega_\mathrm{K}$ of the dense cores is twice as large as the median value of $E_\mathrm{rot}/\Omega_\mathrm{K}$ of the droplets.

As in the case of the size-$\beta$ distribution, a large scatter in the size-$E_\mathrm{rot}/\Omega_\mathrm{K}$ distribution prevents a conclusion.  We note that a potential relation between the size and $E_\mathrm{rot}/\Omega_\mathrm{K}$ would indicate a deviation from an observed velocity pattern that is the result of a turbulence scaling law.  As \citet{Burkert_2000} pointed out, $\omega \propto \sigma/R$ for such structures \citep[see Equation 14 in][]{Burkert_2000}, which would make $E_\mathrm{rot}/\Omega_\mathrm{K} \propto (\omega R/\sigma)^2$ a constant with respect to the size.  A non-constant relation between the size and $E_\mathrm{rot}/\Omega_\mathrm{K}$, if there is one, would suggest that the observed velocity gradient is not fully the result of internal turbulence within these cores.  See Fig.\ \ref{fig:ratioRotKin}.

% ratioRotKin-size and distribution

%%%%

\begin{figure}[ht!]
\plotone{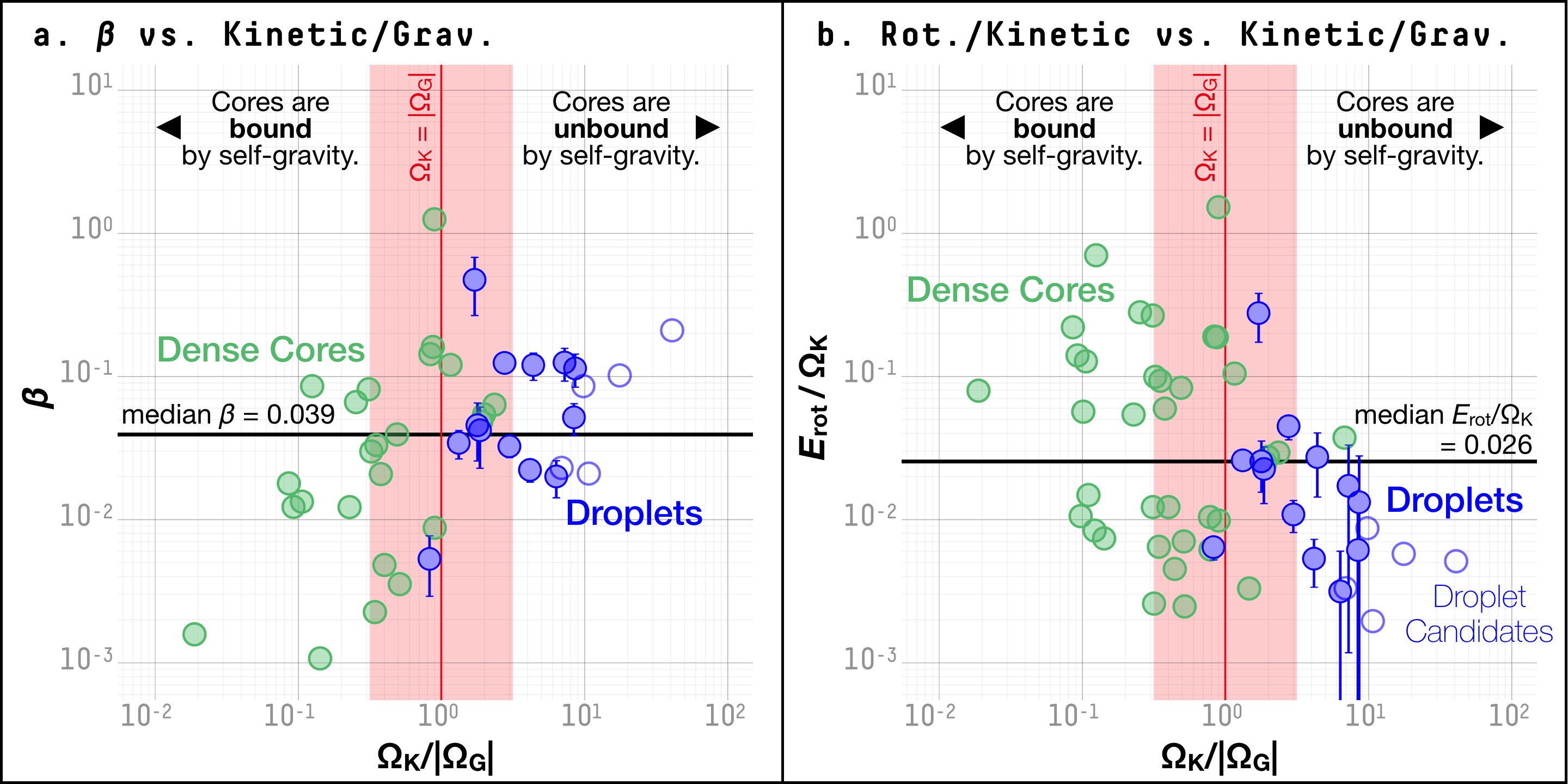}
\caption{\label{fig:betavir} \textbf{(a)} Ratio between rotational and gravitational energies, $\beta$, plotted against the ratio between the kinetic and gravitational energies, $\Omega_\mathrm{K}/\left|\Omega_\mathrm{G}\right|$, for larger-scale dense cores (green dots), droplets (blue filled dots), and droplet candidates (blue empty dots).  The red line corresponds to the relation, $E_\mathrm{rot} = \left|\Omega_\mathrm{G}\right|$, and the red band marks the parameter space within an order of magnitude from this relation.  The parameter space to the right of the red line corresponds $\left|\Omega_\mathrm{G}\right| < \Omega_\mathrm{K}$, and the left corresponds to $\left|\Omega_\mathrm{G}\right| > \Omega_\mathrm{K}$.  The horizontal black line marks the median value of $\beta$ for all cores shown in the figure (both dense cores and droplets; excluding droplet candidates).  \textbf{(b)} Ratio between rotational and total kinetic energies, $E_\mathrm{rot}/\Omega_\mathrm{K}$, plotted against the ratio between the kinetic and gravitational energies, $\Omega_\mathrm{K}/\left|\Omega_\mathrm{G}\right|$, for larger-scale dense cores (green dots), droplets (blue filled dots), and droplet candidates (blue empty dots).  Same as in (a), the red line corresponds to the relation, $E_\mathrm{rot} = \left|\Omega_\mathrm{G}\right|$ and separates the parameter space into one corresponding to objects being bound by self-gravity (left) and another corresponding to objects not being bound by self-gravity (right).  The horizontal black line marks the median value of $E_\mathrm{rot} = \left|\Omega_\mathrm{G}\right|$ for all cores shown in the figure, excluding droplet candidates.}
\end{figure}

Arguably the most important difference between the dense cores and the droplets is in their gravitational \textit{boundedness}.  While the droplets and the dense cores follow the same mass-size and linewidth-size relations (see Fig.\ 9 in Paper I), the droplets are not bound by self-gravity but are instead bound by the pressure provided by the ambient gas motions  (Paper I).  In Fig.\ \ref{fig:betavir}, we then compare $\beta$ and $E_\mathrm{rot}/\Omega_\mathrm{K}$ to the ratio between the internal kinetic energy and the gravitational potential energy, which characterizes the gravitational boundedness of a core.  Fig.\ \ref{fig:betavir}a suggests that there is a mild, if any, tendency for less virially bound structures (mostly droplets) to have larger values of $\beta$, and again, the large scatter in $\beta$ prevents a statistical conclusion of any relation between $\beta$ and the gravitational boundedness.  On the other hand, Fig\ \ref{fig:betavir}b suggests that objects that are more bound by self-gravity tend to have larger $E_\mathrm{rot}/\Omega_\mathrm{K}$ and those less bound by self-gravity have smaller $E_\mathrm{rot}/\Omega_\mathrm{K}$.  In \S\ref{sec:discussion_physics}, we discuss the origin of a potential correlation between gravitational boundedness and the ratio between rotational, kinetic, and gravitational potential energies.

%%%%%
%%%%%
\subsection{Pixel-by-pixel Integration of Angular Momentum}
\label{sec:analysis_pixbypix}
Since Paper I finds that droplets generally have shallow yet not uniform density profiles (see Fig.\ 12 in Paper I), we test the idea of using the observed column density map, instead of assuming a uniform density, when calculating the angular momentum.  To derive the total angular momentum of a droplet, we start by calculating the angular momentum corresponding to each pixel on the observed maps, assuming that at each pixel, the observed mass (column density observed at the pixel, multiplied by the pixel area in physical units) is rotating around the axis of rotation at a velocity equal to the fitted line-of-sight velocity at the pixel (from the linear velocity fits presented in \S\ref{sec:analysis_gradient}).  We also assume that the axis of rotation is in the plane of the sky (such that $\sin{i} = 1$, where $i$ is the inclination angle) with a position angle corresponding to the direction perpendicular to the fitted velocity gradient.  The angular momentum for the $i$-th pixel is then:

\begin{equation}
\label{eq:pixJ}
J_i = m_\mathrm{ave} (N_i-N_\mathrm{min}) V_{\mathrm{fit,}i} R_i \cos{(\theta_\mathcal{G}-\theta_i)}\ \mathrm{,}
\end{equation}

\noindent where $m_\mathrm{ave}$ is the average particle mass \citep[2.8 u;][]{Kauffmann_2008}; $N_i$ is the column density at the $i$-th pixel; $N_\mathrm{min}$ is the minimum column density within the droplet boundary; $V_{\mathrm{fit,}i}$ is the fitted line-of-sight velocity at the $i$-th pixel; $R_i$ is the physical distance in the plane of the sky between the $i$-th pixel and the droplet center; and $\theta_i$ is the position angle of a line in the plane of the sky connecting the $i$-th pixel and the center of the droplet.  Note that we adopt the ``clipping'' paradigm introduced by \citet{Rosolowsky_2008b} and used in Paper I to derive the mass of a droplet, with the minimum column density within the droplet boundary, $N_\mathrm{min}$, as a baseline.  This is done as a way to remove the foreground/background contribution to column density measurements.  Note that this approach is relatively conservative and may lead to an underestimation of the mass.  A similar method is adopted by \citet{Pineda_2015} to estimate the masses of the density structures found inside the coherent core in B5.

The total angular momentum of the droplet is the sum of $J_i$, over all pixels inside the droplet boundary:

\begin{equation}
\label{eq:pixJsum}
J_\mathrm{tot} = \sum_{i\ \mathrm{inside}}{J_i}\ \mathrm{,}
\end{equation}

\noindent and the specific angular momentum of each droplet is $J_\mathrm{tot}/M$, where $M$ is the mass of the droplet (Paper I).  The definition is same as that used by \citet{Rosolowsky_2003}, in an attempt to measure the alignment between the rotational axis and the axis of galaxies \citep[see the definition of $j_2$ in][]{Rosolowsky_2003}.

\begin{figure}[ht!]
\epsscale{0.6}
\plotone{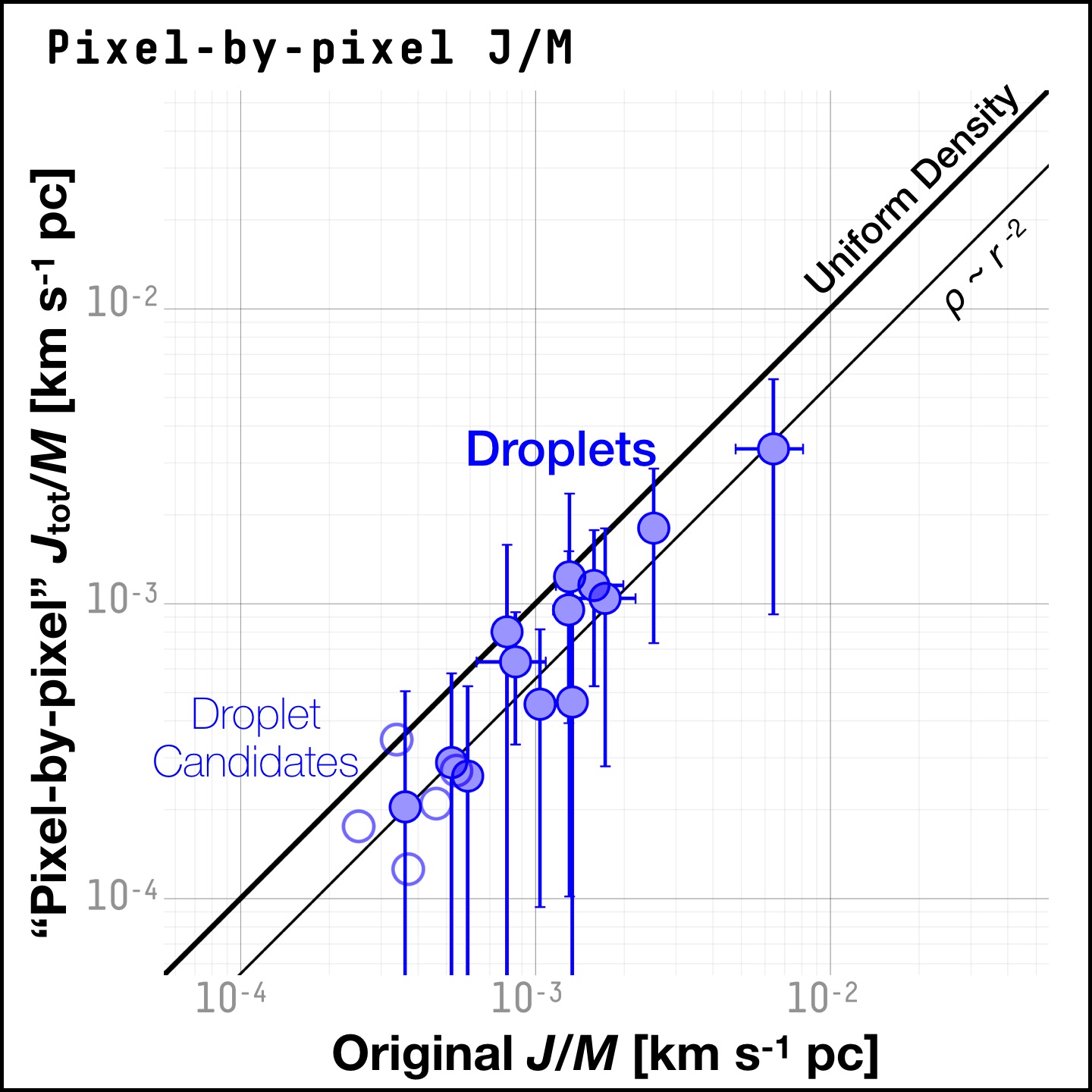}
\caption{\label{fig:JoverM_JoverM} Specific angular momentum derived from pixel-by-pixel integration (``pixel-by-pixel'' $J_\mathrm{tot}/M$), plotted against the specific angular momentum derived in \S\ref{sec:analysis_JoverM}, the latter of which is derived using the method adopted by G93 and assuming solid-body rotation and a uniform density.  The thick line corresponds to when the pixel-by-pixel $J_\mathrm{tot}/M$ is equal to the original $J/M$, i.e.\ when the observed density distribution is consistent with a uniform density.  The thin line corresponds to the expected distribution between the two values of $J/M$ when the observed density follows a power-law distribution with an index of $p = 2$, i.e.\ $\rho \propto r^{-2}$.  See \S\ref{sec:analysis_pixbypix} for details on the derivation of pixel-by-pixel $J_\mathrm{tot}/M$.}
\end{figure}

Fig.\ \ref{fig:JoverM_JoverM} shows the resulting ``pixel-by-pixel'' $J_\mathrm{tot}/M$ compared to the original $J/M$, calculated in \S\ref{sec:analysis_JoverM} assuming a uniform density for each droplet.  The centrally concentrated density profiles of droplets make $J_\mathrm{tot}/M$ generally smaller than $J/M$, since assuming a uniform density is equivalently overestimating the proportion of mass at larger radii.  The difference between the ``pixel-by-pixel'' $J_\mathrm{tot}/M$ and the original $J/M$ is consistent with power-law density profiles with indices between $\sim -2$ and $0$, consistent with what Paper I finds for radial density profiles based on the \textit{Herschel} column density map.

% J/M-J/M

\begin{figure}[ht!]
\plotone{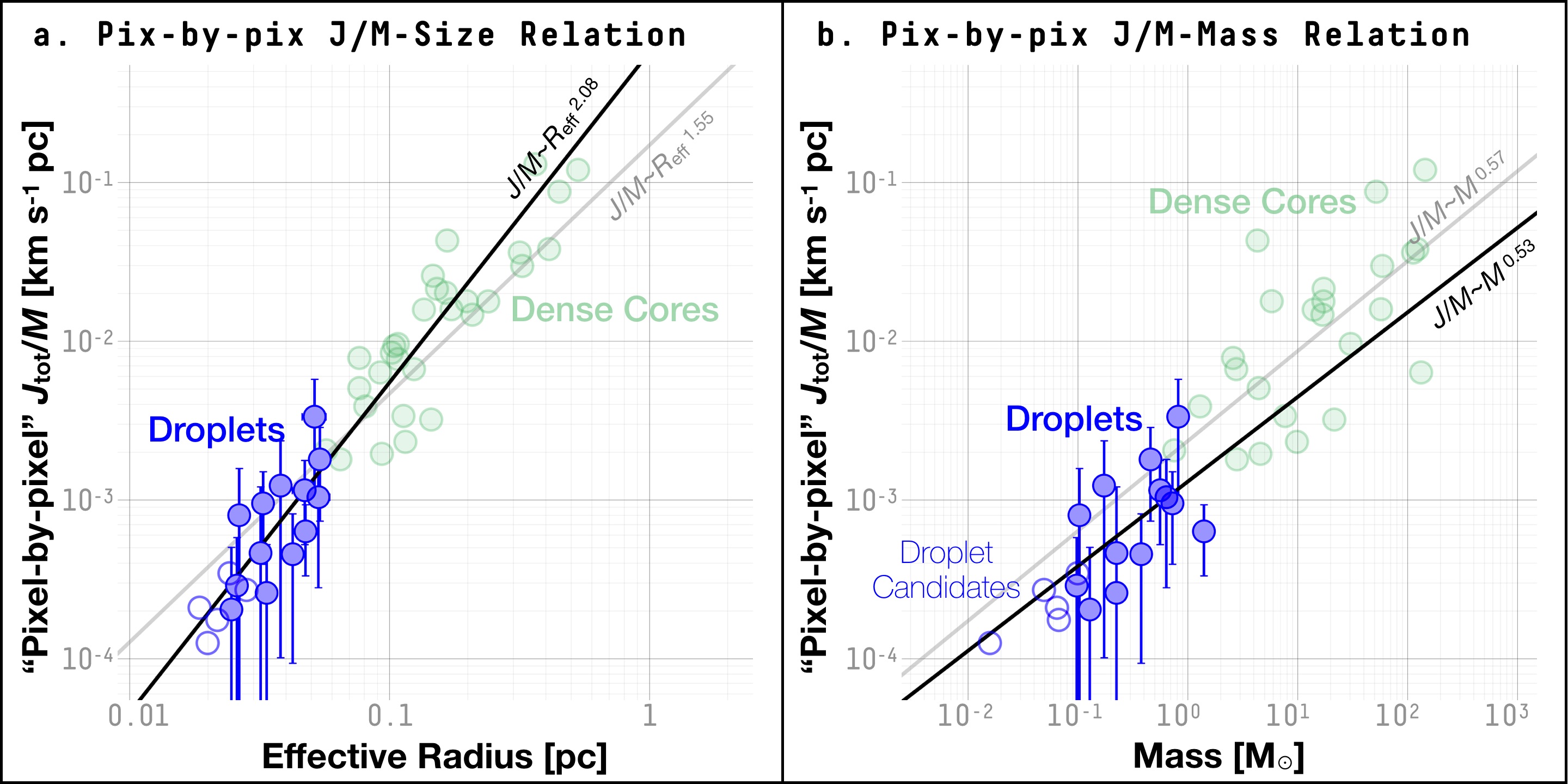}
\caption{\label{fig:JoverMalt} \textbf{(a)} Specific angular momentum derived from pixel-by-pixel integration (``pixel-by-pixel'' $J_\mathrm{tot}/M$), plotted against the effective radius, for droplets (filled blue dots) and droplet candidates (empty blue dots).  Since we do not have the access to the original observations used by G93, the pixel-by-pixel integration of $J/M$ described in \S\ref{sec:analysis_pixbypix} is not applicable to the larger-scale dense cores included in G93.  As a reference, the light green dots are plotted showing the distribution of the \emph{original} $J/M$ and the effective radius for the larger-scale dense cores.   (The light green dots are the same as the green dots shown in Fig.\ \ref{fig:JoverM}a.)  The black line shows a power-law relation between the pixel-by-pixel $J_\mathrm{tot}/M$ and the effective radius, fitted for droplets (excluding the droplet candidates), and the gray line shows the power-law relation between the original $J/M$ and the effective radius found for all cores (the same power-law relation shown in Fig.\ \ref{fig:JoverM}a).  \textbf{(b)} Specific angular momentum derived from pixel-by-pixel integration (pixel-by-pixel $J_\mathrm{tot}/M$), plotted against the mass, for droplets (filled blue dots) and droplet candidates (empty blue dots).  Same as in (a), the light green dots are plotted showing the distribution of the \emph{original} $J/M$ and the mass for the larger-scale dense cores as a reference.   (The light green dots are the same as the green dots shown in Fig.\ \ref{fig:JoverM}b.)  The black line shows a power-law relation between the pixel-by-pixel $J_\mathrm{tot}/M$ and the mass, fitted for droplets (excluding the droplet candidates), and the gray line shows the power-law relation between the original $J/M$ and the effective radius found for all cores (the same power-law relation shown in Fig.\ \ref{fig:JoverM}b).}
\end{figure}

Fig.\ \ref{fig:JoverMalt} plots the ``pixel-by-pixel'' $J_\mathrm{tot}/M$ against the sizes and masses of the droplets.  The $J_\mathrm{tot}/M$-$R_\mathrm{eff}$ relation is steeper than what is found with the uniform density assumption (Fig.\ \ref{fig:JoverM}).  Another interesting aspect of using the ``pixel-by-pixel'' method described in Equation \ref{eq:pixJsum} is that it does not assume a rotational profile, such as the solid-body rotation assumed in the original $J_\mathrm{tot}/M$ measurements in \S\ref{sec:analysis_JoverM} and in G93.  Intriguingly, the resulting ``pixel-by-pixel'' $J_\mathrm{tot}/M$-size relation for the droplets is consistent with solid-body rotation with a power-law index close to 2 (Fig.\ \ref{fig:JoverMalt}).  Due to the relatively small range of size and large uncertainties in the ``pixel-by-pixel'' measurement of $J_\mathrm{tot}/M$, the $J_\mathrm{tot}/M$-size relation shown in Fig.\ \ref{fig:JoverMalt} can also be consistent with the rotational profile, $j(r)$ $\propto$ $r^{1.8}$, observed by \citet{Pineda_2019} using interferometric observations, which indicates that the observed density distributions are likely results of turbulence and solid-body rotation at the same time.  (See discussions below in \S\ref{sec:discussion_physics}.)  Unfortunately, since we do not have access to the original observations presented by G93, we cannot derive the specific angular momenta for larger-scale dense cores with the above method (Equation \ref{eq:pixJsum}) nor a ``pixel-by-pixel'' $J_\mathrm{tot}/M$-size relation over a larger range of size.

The overall smaller ``pixel-by-pixel'' $J_\mathrm{tot}/M$ and a steeper $J_\mathrm{tot}/M$-$R_\mathrm{eff}$ relation would make the measured rotational energy, $E_\mathrm{rot}$, shown in Fig.\ \ref{fig:Erot} smaller.  This would in turn make the two energy ratios, $\beta$ and $E_\mathrm{rot}/\Omega_\mathrm{K}$, discussed in \S\ref{sec:analysis_beta} smaller for the droplets.  Although the ``pixel-by-pixel'' measurements should not be directly compared to measurements made by G93 using an assumption of constant density and solid-body rotation, the smaller $\beta$ for the droplets would make the $\beta$-size relation even flatter (see Fig.\ \ref{fig:beta}).  For the $E_\mathrm{rot}/\Omega_\mathrm{K}$-size relation, smaller $E_\mathrm{rot}/\Omega_\mathrm{K}$ for the droplets would make the potential correlation between $E_\mathrm{rot}/\Omega_\mathrm{K}$ and size more significant (see Fig.\ \ref{fig:ratioRotKin}).  In summary, using $J_\mathrm{tot}/M$ measured using the ``pixel-by-pixel'' method would not qualitatively change the results presented in \S\ref{sec:analysis_beta}.  Below in \S\ref{sec:discussion_physics}, we discuss the implication of ``pixel-by-pixel'' measurements and compare it to measurements of rotational profiles made with higher-resolution interferometric observations.

% J/M alt - mass/size

\subsection{Alignment between the Velocity Gradient and the Core Shape}
\label{sec:analysis_shape}
To examine the alignment between the velocity gradient and the core elongation, we adapt the methods introduced by \citet{Soler_2013} and applied on Planck surveys of column density and dust polarization by \citet{Planck_35}.  \citet{Planck_35} used the histogram of relative orientations (HRO) to analyze the alignment between the magnetic field direction, traced by Planck observations of polarized dust emission, and the column density gradient which measures the orientations of column density structures.  \citet{Planck_35} used the HRO to show the distribution of the difference in position angles between two sets of vectors in the plane of the sky, with each vector measured at a pixel in the survey map.  The HRO was applied on cloud-scale structures, much larger in size compared to the coherent structures in L1688 and B18.  Here we adapt the HRO for the samples of individual objects (droplets and larger-scale dense cores) and for tracing the difference between the velocity gradient orientation and the position angle of the major axis, the latter of which is derived using a principal component analysis (PCA) of the NH$_3$ brightness distribution and defined between $-90$ and $90$ degrees (East of North; see Table \ref{table:gradient}).

\begin{figure}[ht!]
\plotone{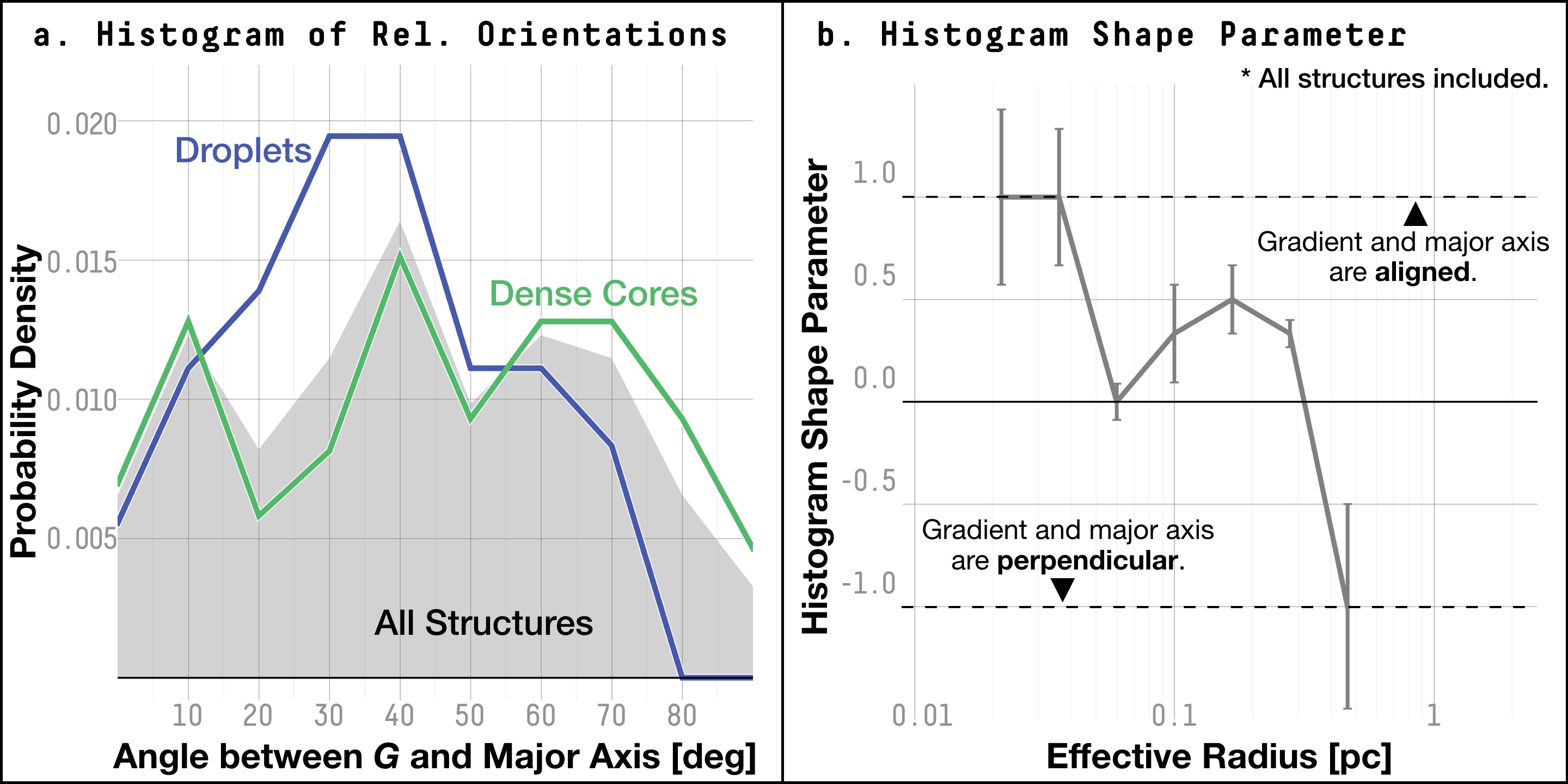}
\caption{\label{fig:HRO} \textbf{(a)} Histogram of relative orientations (HRO) between the velocity gradient orientation and the position angle of the major axis, for larger-scale dense cores (the green line), droplets (the blue line), and both populations combined (the gray histogram).  The normalized probability density is derived from a kernel density estimation (KDE) analysis.  The HROs shown in this figure do not distinguish between angle differences in clockwise and counterclockwise directions, either.  (That is, the distributions of relative orientations shown in \citet{Soler_2013} going from $-90$ degrees to $90$ degrees would be ``folded'' along 0 degree if plotted in this figure.)  \textbf{(b)} ``Histogram shape parameter'' for the HRO shown in (a), measured at different values of the effective radius, using both the larger-scale dense cores and the droplets.  An ideal case where the velocity gradient aligns with the major axis in every core would generate a value of $1$, and a case when the velocity gradient is perpendicular to the major axis in every core would generate a value of $-1$.  The definition of the ``histogram shape parameter'' is modified for the samples of individual objects examined in this work.  See \S\ref{sec:analysis_shape} for details.}
\end{figure}

The angle between the velocity gradient orientation and the position angle of the major axis is defined to be the smaller angle spanned by the gradient vector and the major axis and is defined between 0 and 90 degrees.  We do not distinguish between the position angle differences in clockwise and counterclockwise directions (see Fig.\ \ref{fig:HRO}a).  That is, a distribution of relative orientations shown in \citet{Planck_35} going from $-90$ degrees to $90$ degrees would be ``folded'' along 0 degree.  Fig.\ \ref{fig:HRO}a shows the histogram of relative orientations (HRO) for the droplets, the larger-scale dense cores, and both groups combined.
% And, instead of the normalized distribution density, we plot the frequency of objects in each population falling in each histogram bin, mainly because of the relatively small number of samples (13 droplets and 24 dense cores).  

Judging from the HROs shown in Fig.\ \ref{fig:HRO}a, there is no clear sign of preference in the alignment between the velocity gradient and the major axis of the core.  To quantify the result, we modify the histogram shape parameter introduced by \citet{Soler_2013} to directly use the numbers of samples within certain ranges of relative orientations, instead of the areas within certain angle ranges on the HRO.  The adapted ``histogram shape parameter'' (now being independent of the histogram) is

\begin{equation}
\label{eq:HSP}
\xi = \frac{N_\mathrm{c} - N_\mathrm{e}}{N_\mathrm{c} + N_\mathrm{e}}\ \mathrm{,}
\end{equation}

\noindent where $N_\mathrm{c}$ is the number of cores with relative orientations smaller than 22.5 degrees, and $N_\mathrm{e}$ is the number of cores with relative orientations larger than 67.5 degrees.  The choice of angle ranges is the same as that used by \citet{Soler_2013} and \citet{Planck_35}.  For a population where the velocity gradient aligns with the major axis of each object, $\xi = 1$, and for a population where the velocity gradient is perpendicular to the major axis of each object, $\xi = -1$.

For the entire sample of droplets and dense cores, we measure $\xi = 0.25$, with $N_\mathrm{c} = 15$ and $N_\mathrm{e} = 9$.  Fig.\ \ref{fig:HRO}b shows the ``histogram shape parameter'' measured for objects with different sizes.  We find that $\sim$ 44\% of the droplets with significant detections of linear velocity gradients have angle differences between $\theta_\mathrm{G}$ and $\theta_\mathrm{shape}$ smaller than 22.5 degrees (e.g.\ L1688-d4, L1688-d7 and B18-d4; see Figs.\ \ref{fig:gallery1} to \ref{fig:gallery3}), while the rest have angle differences between 22.5 and 67.5 degrees.  None of the droplets have angle differences larger than 67.5 degrees (Fig.\ \ref{fig:HRO}a).  Overall, while there seems to be a tendency for smaller/larger cores to have velocity gradients parallel/perpendicular to the elongations of the cores at first glance (Fig.\ \ref{fig:HRO}b), we note that the small number of cores can bias the result.

Considering that the larger-scale cores may be older, as suggested by their gravitational boundedness (Paper I), the results that there is potentially a tendency for them to have elongation perpendicular to the local velocity gradient can be consistent with the results from observations of dense cores in Perseus.  Using observations of N$_2$H$^+$ emission, Che-Yu Chen et al.\ (\textit{submitted}) find that the disk structures are perpendicular to the local velocity gradients in a Class 0/I object (Per 30) and two starless cores (B1-NE and B1-SW; see Table 3 in Che-Yu Chen et al.).  These results are opposite to what the classical theory of disk formation describes, wherein the disk and its rotational motion is directly inherited from the initial angular momentum in the core obtained via accretion of materials at larger distances.  On the other hand, while Che-Yu Chen et al.\ (\textit{submitted}) find that the observed velocity gradients are likely a continuation of velocity structures at larger scales and thus likely originate in the larger-scale turbulence and/or convergent flow compression, we find that in many cases analyzed in this paper, the velocity structures of droplets are disconnected from the surrounding regions (for example, see L1688-d5 and L1688-d6 in Fig.\ \ref{fig:gallery1} and B18-d4 in Fig.\ \ref{fig:gallery3}).

% HRO
%The histogram of relative orientations (HRO) between the velocity gradient orientation and the position angle of the major axis, for larger-scale dense cores (the green line), droplets (the blue line), and both populations combined (the gray histogram).  Different from HROs shown in \citet{Soler_2013, Planck_35}, the frequencies of objects in each group falling in the histogram bins are plotted along the y-axis, instead of a normalized distribution density (see \S\ref{sec:discussion_shape}).  The HROs shown in this figure do not distinguish between angle differences in clockwise and counterclockwise directions, either.  (That is, the distributions of relative orientations shown in \citet{Soler_2013} going from $-90$ degrees to $90$ degrees would be ``folded'' along 0 degree if plotted in this figure.)

%%%%%
%%%%%

\section{Discussion: Physical Interpretation}
\label{sec:discussion_physics}
% do I need to summarize here?  at least include some guidelines for what I'm trying to do in this section.
%In \S\ref{sec:analysis}, we derive the velocity gradients for droplets, and find that the droplets generally follow the same power-law relation between the velocity gradient and the size/mass found for larger-scale dense cores \citep{Goodman_1993}.  Assuming that the fitted velocity gradient arises from a solid-body rotation of a spherical rotating body with a uniform density, we derive the specific angular momentum, $J/M$, and the ratio between the rotational and gravitational energies, $\beta$.

%\subsection{Physical Interpretation}
%\label{sec:discussion_physics}
% Summary:
% 1. We find $\beta = 0.05_{-0.02}^{+0.08}$, roughly a factor of 1.5 larger than 0.032 found by G93 for the dense cores.  Overall, we find $\beta$ $\sim$ 0.04 for both populations with a large scatter (Fig.\ \ref{fig:Erot}a and Fig.\ \ref{fig:beta}).
% 2. The ratio between the rotational energy and the total kinetic energy of the larger-scale dense cores appears to be generally larger than the droplets (Fig.\ \ref{fig:Erot}b).
% 3. We find that the relation between the specific angular momentum and the size extends to smaller scales, following a power-law with an index of 1.55.

% need references: Burkert_2000, Li_2004, CChen_2018, Pineda_2019, magnetic braking (see references in Offner_2008)

% beta
For the droplets, we find a typical value of $\beta = 0.046_{-0.024}^{+0.079}$, roughly a factor of 1.5 larger than 0.032 found by G93 for the dense cores.  Overall, we find $\beta$ $\sim$ 0.039 that is generally independent of sizes with a larger scatter in $\beta$ for both droplets and dense cores (Fig.\ \ref{fig:Erot}a and Fig.\ \ref{fig:beta}).  In comparison, \citet{Offner_2008} find $\beta$ $\sim$ 0.05 for star-forming bound ``clumps'' in a gravoturbulent simulation with driven turbulence and $\beta$ $\sim$ 0.08 in a gravoturbulent simulation with decaying turbulence.  \citet{Dib_2010} find smaller $\beta$ ranging from $\sim$ 0.014 to $\sim$ 0.026 in a survey of magnetized and self-gravitating dense cores in simulations of magnetized molecular clouds with decaying turbulence.  The observed $\beta$ for the droplets seems to agree best with the value measured for bound cores in simulations with driven turbulence in \citet{Offner_2008}.  It is worth noticing that in \citet{Offner_2008}, simulations with driven turbulence also produced cores with subsonic velocity dispersions while simulations with decaying turbulence did not \citep[see Fig.\ 3 in][]{Offner_2008}.  \citet{Offner_2008} suggested that magnetic braking may serve as an efficient means to further suppress $\beta$ by transferring angular momentum outwards \citep[see also][]{Hosking_2004, Banerjee_2006}.  The process of magnetic braking, especially at smaller scales of $\lesssim$ 1000 AU, may partly explain the slightly smaller value of $\beta$ found for larger-scale dense cores (via averaging over entire cores), since the magnetic field strengths in these denser cores is expected to be larger than in the gravitationally unbound droplets.  The larger-scale cores may also be older, as suggested by their gravitational boundedness, giving more time for the process of magnetic braking to act.  \citet{Li_2004} find a similar value of $\beta$ $\sim$ 0.05 in an ideal MHD simulation, although the cores are supercritical by an order of magnitude.  A study of both gravitationally bound and pressure bound coherent structures in simulations is needed to further examine the physical processes involved in the evolution of rotational motions in droplets and coherent cores.

% gravity-organized
Considering that the larger-scale dense cores may be older as suggested by Paper I, the relation between the rotational, kinetic, and gravitational potential energies, shown in Fig.\ \ref{fig:betavir}, may be consistent with a picture where the gravitational infall provides the initial angular momentum and the rotational motions co-evolve with gravity.  A non-constant size-$E_\mathrm{rot}/\Omega_\mathrm{K}$ relation, if there is one (see Fig.\ \ref{fig:ratioRotKin}), would also indicate that the measured velocity gradient is not fully the result of turbulence within cores.  On the other hand, \citet{Burkert_2000} find that the observed velocity pattern can be the result of a turbulence scaling law.  For turbulence dominated gas with a Larson's law-like power-law relation between velocity dispersion and the size, $\delta v$ $\sim$ $r^{0.5}$, there exists a relation between the observed specific angular momentum and the size, $J/M$ $\propto$ $R^{1.5}$, in the model presented by \citet{Burkert_2000}.  \citet{CChen_2018} also find $J/M$ $\propto$ $R^{1.5}$ in a survey of gravitationally bound dense cores in MHD simulations and conclude that the rotational motion is acquired from ambient turbulence.  Similarly, Che-Yu Chen et al.\ (\textit{submitted}) conclude that the observed velocity structures likely originate in the large-scale turbulence or convergent flow, using observations of N$_2$H$^+$ of dense cores in Perseus.

Using Very Large Array (VLA) interferometric observations of NH$_3$ (1, 1) emission, \citet{Pineda_2019} examine the interior velocity structures in two Class 0 objects and one first hydrostatic core candidate.  Resolving the internal velocity structures within these cores, \citet{Pineda_2019} are able to derive the radial profile of specific angular momentum, $j(r)$, and find $j(r)$ $\propto$ $r^{1.8}$.  The result suggests that the observed ``net rotational motion'' in cores is not purely solid-body rotation as usually assumed, which would result in $j(r)$ $\propto$ $r^{2}$, and the turbulence is involved in creating the observed velocity structures in cores.  In this study, we derive a $J/M$-size relation consistent with that derived by G93, $J/M$ $\propto$ $R^{1.5}$, assuming that the observed cores have constant densities and the velocity structure is due to solid-body rotation (see \S\ref{sec:analysis_JoverM}).  Using column density maps derived from Herschel observations, we also derive a ``pixel-by-pixel'' $J_\mathrm{tot}/M$-size relation, $J_\mathrm{tot}/M$ $\propto$ $R^{2.08}$ for the droplets over a limited range of sizes (see \S\ref{sec:analysis_pixbypix}), without having to assume either a constant density or solid-body rotation.  Again, as G93 pointed out, measurements of rotational quantities based on observations represent a ``net rotational motion'' instead of indicating that the core is actually rotating in a solid-body manner.  The measurement is also likely affected by observational effects such as the uncertainty in the viewing angle.  More studies need to be done on the co-evolution between gravity, turbulence, the magnetic field, and the rotational motion.  In particular, it could benefit from analyses of velocity gradients in the interiors of dense cores found in higher-resolution observations of higher-density tracers, such as those observed by the MASSES survey \citep{Stephens_2018}.

\section{Conclusion}
\label{sec:conclusion}
In this paper, we use the data from Green Bank Ammonia Survey \citep[GAS;][]{GAS_DR1} and examine the internal velocity structures of the droplets---sub-0.1 coherent core-like structures, in two nearby star forming regions, L1688 in Ophiuchus and B18 in Taurus (Paper I).  A linear velocity field is fitted to the observed $V_\mathrm{LSR}$, derived from NH$_3$ hyperfine line fitting (Fig.\ \ref{fig:gallery1}, \ref{fig:gallery2}, and \ref{fig:gallery3}).  The resulting velocity gradients of the droplets are found to follow a power-law relation between the gradient magnitude and the size similar to the relation found for larger-scale dense cores by G93, although with some dispersion in the fitted velocity gradient (Fig.\ \ref{fig:Gradient}).

Following G93, we assume that the fitted velocity gradient arises from solid-body rotation of a rotating sphere with a uniform density.  We derive the specific angular momentum, $J/M$, and find that both the droplets and the larger-scale dense cores appear to follow a relatively tight relation between $J/M$ and the effective radius (Fig.\ \ref{fig:JoverM}a).  The relation between $J/M$ and the size found for droplets is consistent with what G93 find for larger-scale dense cores, $J/M$ $\propto$ $R^{1.5}$.  However, we note that the tight correlation is at least partly due to the potentially wrong assumptions of solid-body rotation and a uniform density for the observed core.  Also, as numerous works on simulations and observations have pointed out, a $J/M$-size relation of $J/M$ $\propto$ $R^{1.5}$ is consistent with turbulent motions that follow a Larson's law-like scaling relation \citep[e.g.][]{Burkert_2000, CChen_2018}.

We examine the rotational energy of the assumed solid-body rotation that gives rise to the fitted velocity gradient.  We find that the rotational energy, $E_\mathrm{rot}$, is generally smaller than the gravitational energy, $\Omega_\mathrm{G}$ by an order of magnitude or more (Fig.\ \ref{fig:Erot}a), which suggests that self-gravity alone can provide the needed binding to sustain the rotational motion.  We also compare $E_\mathrm{rot}$ to the internal kinetic energy calculated for the thermal and non-thermal motions of gas inside the droplet/larger-scale dense core, $\Omega_\mathrm{K}$ (Fig.\ \ref{fig:Erot}b).  We find $E_\mathrm{rot}$ is generally smaller than $\Omega_\mathrm{K}$, suggesting that the rotational contribution to the observed linewidth is smaller than the contribution from thermal and turbulent motions inside droplets and larger-scale dense cores.

For the droplets, we find a typical value of the ratio between the rotational energy and the gravitational potential energy, $\beta$, of $0.046_{-0.024}^{+0.079}$.  Overall, for the entire population including both the droplets and the larger-scale dense cores examined in G93, we find $\beta \sim 0.039$.  Consistent with what G93 found for larger-scale dense cores, we find that $\beta$ generally independent of the size scale for the droplets.  The result extends the scale independency of $\beta$ down to a size scale of $\sim$ 0.02 pc.  As G93 pointed out, this can be due to the fact that NH$_3$ emission traces almost constant-density gas.  Similarly, there is a large scatter in the distribution of $E_\mathrm{rot}/\Omega_K$.

%On the other hand, we find that the ratio between the rotational energy and the total kinetic energy, $E_\mathrm{rot}/\Omega_K$, potentially has a tendency of increasing towards larger-scale, more gravitationally bound cores.  Considering that the larger-scale dense cores may be older as suggested by Paper I, the result is consistent with a picture where the rotational motion comes from the angular momentum obtained via accretion and co-evolves with gravitational infall.

We also derive the specific angular momentum directly from the \textit{Herschel} column density maps, instead of having to assume a uniform density and solid-body rotation for each droplet.  Consistent with what Paper I finds for the radial density profiles, the difference between the ``pixel-by-pixel'' specific angular momentum and the original value is consistent with power-law density profiles with indices between $-2$ and $0$ (Fig.\ \ref{fig:JoverM_JoverM}).  Consequently, with the ``pixel-by-pixel'' specific angular momentum, we find a steeper relation between the specific angular momentum and the size (Fig.\ \ref{fig:JoverMalt}).

Adapting the methods used by \citet{Soler_2013} and \citet{Planck_35} to quantify the alignment between the polarization and the orientation of column density structures, we show that there is barely any indication of preferred alignment between the velocity gradient and the major axis of a core included in this work (Fig.\ \ref{fig:HRO}a).  The adapted ``histogram shape parameter'' seems to suggest that there could be a tendency for smaller cores to have velocity gradients parallel to the major axes.  However, the small number of samples prevents any statistical conclusions.

As shown in this study, the observations of velocity gradients in droplets not only provide an observational constraint on the measurement of ``net rotational motions'' at size scales $\lesssim$ 0.1 pc, they may also provide a look into the velocity structures at a potentially earlier stage in core evolution.  The results presented in this paper suggest a tight relation between gravity, turbulence, and the rotational motion.  More research needs to be done to study the co-evolution of rotation with other physical processes in the process of core formation.

%\doi{10.5281/zenodo.15991}  

\acknowledgments
This work was supported by a Cottrell Scholar Award.  The National Radio Astronomy Observatory is a facility of the National Science Foundation operated under cooperative agreement by Associated Universities, Inc.
%This research made use of Astropy, a community-developed core Python package for Astronomy \citep{astropy}.

\vspace{5mm}
\facilities{GBT (KFPA+VEGAS), Herschel Space Observatory (PACS+SPIRE)}

\software{astropy \citep{astropy}, Glue \citep{glue, glue_2017}, PySpecKit \citep{pyspeckit}, RADMC-3D \citep{radmc3d}}

\bibliography{main.bib}

\end{document}